\documentclass{article}

\usepackage{PRIMEarxiv}

\usepackage[utf8]{inputenc} 
\usepackage[T1]{fontenc}    
\usepackage{hyperref}       
\usepackage{url}            
\usepackage{booktabs}       
\usepackage{amsfonts}       
\usepackage{nicefrac}       
\usepackage{microtype}      
\usepackage{lipsum}
\usepackage{fancyhdr}       
\usepackage{graphicx}       
\graphicspath{{media/}}     

\usepackage{multirow}
\usepackage{multicol}
\usepackage{rotating}
\usepackage{algorithm, algorithmicx, algpseudocode}
\usepackage{listings}%
\usepackage{alltt}%
\usepackage{amsmath}
\usepackage{amssymb}
\usepackage{dcolumn}

\title{Prediction-based Variable Selection for Component-wise Gradient Boosting
}

\author{
  Sophie Potts, Elisabeth Bergherr, Colin Griesbach \\
  Chair of Spatial Data Science and Statistical Learning \\
  University of Goettingen \\
  Goettingen, Germany\\
  \texttt{\{Sophie Potts\}sophie.potts@uni-goettingen.de} \\
   \And
  Constantin Reinke \\
  Chair of Empirical Methods in Social Science and Demography \\
  University of Rostock \\
  Rostock, Germany\\
}

\begin{document}
\maketitle

\begin{abstract}
    \noindent Model-based component-wise gradient boosting is a popular tool for data-driven variable selection. In order to improve its prediction and selection qualities even further, several modifications of the original algorithm have been developed, that mainly focus on different stopping criteria, leaving the actual variable selection mechanism untouched. We investigate different prediction-based mechanisms for the variable selection step in model-based component-wise gradient boosting. These approaches include Akaikes Information Criterion (AIC) as well as a selection rule relying on the component-wise test error computed via cross-validation. We implemented the AIC and cross-validation routines for Generalized Linear Models and evaluated them regarding their variable selection properties and predictive performance. An extensive simulation study revealed improved selection properties whereas the prediction error could be lowered in a real world application with age-standardized COVID-19 incidence rates.
\end{abstract}


\section{Introduction}\label{sec1}

In regression settings with a large number of possible covariates, tools for proper variable selection are useful. One popular tool for data-driven variable selection are model-based component-wise gradient boosting algorithms \cite{Buhlmann.2007}, which emerged in the statistical learning community within the last two decades.

The main idea of the first boosting algorithm {AdaBoost} \cite{Freund.1996} was to fit several simple classifiers and combine these ``weak learners" to a single strong learner with a very good predictive performance. Boosting is therefore also referred to as an ensemble learning technique. Based on the observation that {AdaBoost} minimizes some loss function \cite{breiman1997arcing}, it has been shown that boosting can be represented in the statistical framework of Generalized Additive Models (GAMs) \cite{Friedman.2000}. This allows for more flexibility in terms of loss functions and weak learners such as ordinary-least squares (OLS) fits or splines. Combining boosting with gradient descent techniques to iteratively minimize loss functions made the boosting approaches even more powerful \cite{Friedman.2001}. 
Despite its implicit variable selection, model-based component-wise gradient boosting algorithm as proposed by B\"uhlmann and Hothorn \cite{Buhlmann.2007}, results in a good predictive performance. It combines further advantages like fast computation, coefficient shrinkage and model choice as well as the applicability in high-dimensional settings, where the number of covariates exceeds the number of observations and offers a flexible framework, that can be used for various applications. Since we apply model-based component-wise gradient boosting we refer to it as boosting from now on.

Boosting is a statistical learning mechanism, that iteratively estimates the coefficient vector of a given regression model. To do so, it refits simpler estimation procedures on the (pseudo-)residuals of the previous iteration, which  can be seen as the gradient of a differentiable loss function. In the case of Generalized Linear Models (GLMs), this loss is usually chosen as the negative of the corresponding log-likelihood. For each iteration, the model fit is updated in a component-wise fashion, i.e.\ every single covariate is fitted separately and the algorithm identifies and updates the covariate whose update reduces the loss function the most. 

The boosting procedure is run for a rather large pre-specified number of iterations. Following, for every iteration-specific coefficient vector, prediction criteria can be calculated. This allows to select the best performing coefficient vector in the aftermath. If some covariates did not enter the model up to the selected iteration, they are excluded from the final model. Furthermore, selecting the best iteration controls the bias-variance trade-off and thus enables coefficient shrinkage.

There exist several modifications of the original boosting algorithm, that aim to optimize its performance even further, e.g.\ to lower the false positive rate (FPR) such that only true influential variables enter the final model: %
%
%
The AIC-subsampling approach \cite{Mayr.2012} as well as the probing method \cite{Thomas.2017} focus on a fully data-driven iteration selection.

\noindent Stability selection \cite{meinshausen2010stability, hofner2015controlling}, deselection \cite{stromer2021deselection} and Twin-boost \cite{buhlmann2010twin} concentrate on filtering the main influential covariates by enlarging the number of boosting rounds. Other approaches \cite{staerk2021subspace} allow for multivariate updates and use a double-checking mechanism to decrease the {FPR}.

The majority of approaches mainly focuses on different stopping criteria, leaving the actual \emph{variable selection} mechanism untouched.
Since boosting is a greedy algorithm this variable selection step is of major importance. Thus, in this paper we investigate the variable selection step in order to increase the sparsity of boosted {GLMs} without sacrificing the good predictive performance. 

Only few other papers deal with modifications of the variable selection step. 
One approach modified the variable selection criterion from the traditional loss function to the generalized minimum description-length (gMDL) concept to generate sparser models \cite{buhlmann2006sparse}. When compared to conventional squared-error loss boosting, Sparse$L_2$Boosting with a {gMDL} variable selection produces sparser models with equivalent predictive performance, and is especially effective in high-dimensional settings. 
In the framework of unbiased model selection in boosting with different kinds of covariates, it has been proposed to select the base learner using the {AIC} \cite{Hofner.2011}. However, this does not result in an unbiased model selection in the case of base learners with very different degrees of freedom (e.g. categorical variable with ten categories vs. one metric variable with the same kind of base learner). 
The AIC-based variable selection for likelihood-based boosting has been implemented by Tutz and Groll in the special case of Generalized Linear Mixed Models \cite{tutz2010generalized} and there may be differences in behaviour on {GLMs} owing to the error variance decomposition in mixed models.

There is, however, a lack of a systematic investigation of the modification of the variable selection step in gradient boosted {GLMs} which this paper will address. Therefore, two different prediction-based variable selection mechanisms are presented and evaluated in the following in order to examine their impact on sparsity of gradient boosted {GLMs}. Instead of minimizing a loss function, our approach focuses on minimizing various prediction criteria directly.

The rest of the paper is structured as follows: in the second section, model-based component-wise gradient boosting for {GLMs} is described in more detail and our extension of the concept is explained. In section three a simulation study is conducted to evaluate the new procedures and the results are presented. Section four deals with modelling a real world data set on COVID-19 incidence rates. In section five the results of this paper are summarized and an outlook on possible extensions is given.

\section{Prediction-based variable selection}\label{sec2}

Investigating the variable selection step of boosting is motivated by its direct impact on the variable selection process compared to a more downstream impact of the boosting iteration selection. The selection of the variable to be updated in boosting algorithms usually employs the loss function. Based on the promising findings using the {gMDL}-criterion \cite{buhlmann2006sparse}, this paper investigates two different variable selection criteria for boosting of {GLMs}.

Pulling the former iteration-selection criteria into the variable selection loop of the algorithm may prevent the inclusion of false positives (FPs) already during the fitting process.
Since boosting is a greedy algorithm, i.e.\ once a variable enters the model, it will be included in every following iteration, it is even more important to take a closer look at the variable selection step.
In this section we will explain our modification of this selection step by lining out the changes in the algorithm.

Based on the given data $\mathcal{D} := \{(\boldsymbol {x}_1, y_1),(\boldsymbol {x}_2, y_2),  \ldots, (\boldsymbol {x}_N, y_N) \}$
an iterative algorithm is run. The data can be rewritten as a design matrix $\boldsymbol X \in \mathbb{R}^{N \times (k+1)}$ summarizing all observations $\boldsymbol{x}_i = (x_{i0}, x_{i1}, x_{i2}, \ldots, x_{ik})'$ with $x_{i0}=1$, $i = 1,\dots,N$, $j = 1,\dots,k$  and an outcome vector $\boldsymbol y \in \mathbb{R}^{N}$, with $k \in \mathbb{N}$ being the number of covariates and $N \in \mathbb{N}$ the number of observations.

The algorithm is run for $T$ iterations where each iteration-specific argument is marked by $(\cdot)^{(t)}$ with $t \in \{0, 1,\ldots, T\}$. It is assumed that the regression problem can be written as a {GLM}, where the negative Log-likelihood serves as a loss function for the boosting algorithm in order to estimate the vector of the true coefficients $\boldsymbol \beta \in \mathbb{R}^{k+1}$.

Both upcoming prediction-based variable selection mechanisms possess the advantage, that they do not need further time-consuming computations to determine the final stopping iteration. Since they incorporate a prediction-based measure, which is evaluated for each covariate in every iteration, the global minimum of the $k \times T$ candidates can be used to define the stopping iteration
\begin{eqnarray}
    t^* =     \underset{c ~\in \{1,\ldots, k \times T\}}{\operatorname{argmin}}  \text{crit}_c.
\end{eqnarray}
The selection is done, after the $T$ iterations were run. However, in order to favor sparser models, a minimum improvement threshold of $10^{-8}$ is specified. If the change in the prediction criterion or the $\boldsymbol{\hat \beta}$ vector between two iterations $t$ and $t+1$ is smaller than the threshold, iteration $t$ is selected even though $t+1$ yields the analytical minimum. The value of the threshold is chosen rather arbitrarily to stop fitting if changes become too small. 

\subsection{Cross-validation (CV)}

The first option makes use of an ($F$-fold) CV procedure. The algorithm is essentially identical to boosting, except the variable selection part displayed in Algorithm \ref{algo: cv boost}. It allows for a stable prediction-based evaluation of the best fitting covariate.
Once the optimal covariate-index $j^* \in \{1,\dots,k\}$ is identified, $\boldsymbol {\hat \beta}^{(t)}$ is updated based on the corresponding base learner using all data again.

\begin{algorithm}[ht!]
\begin{algorithmic}
 \vspace{3pt}
 \For {j in $1:k$}
    \For{f in $1:F$}

        \State calculate $j$-th base learner on all but fold $f$ 
        
        \vspace{9pt}
        \setlength\parindent{50pt}
        base learner: ${\hat \beta_{j,-f}^{(t)}}= \left(\left(\boldsymbol{x}_{(j),-f}\right)'\boldsymbol{x}_{(j),-f}\right )^{-1} \left(\boldsymbol{x}_{(j),-f} \right)' \boldsymbol{u}^{(t)}_{-f}$
        
        \vspace{9pt}
        \setlength{\parindent}{0pt}
        \State update $j$-th entry of ${\boldsymbol {\hat \beta}^{(t-1)}}$ by the base learner
        using step length parameter $\nu$ resulting in ${\boldsymbol {\tilde \beta}_{j,-f}^{(t)}}$ 
        
        \State evaluate loss function with ${\boldsymbol {\tilde \beta}_{j,-f}^{(t)}}$ and data from fold $f$ 
        
        \vspace{9pt}
        \setlength\parindent{50pt}
        $l_{j_f}= L \left(\boldsymbol{y}_{{f}}, \boldsymbol{X}_{{f}} {\boldsymbol {\tilde \beta}_{j,-f}^{(t)}}\right)$
        \vspace{9pt}
        
    \EndFor
  \State get mean loss over the folds $\displaystyle{\bar {l_j}= \frac 1 F \sum_{f=1}^F l_{j_f}}$ 
    \EndFor
    \State choose $\displaystyle{j^* = \underset{j ~\in \{1,\ldots,k\}}{\operatorname{argmin}} ~ \bar {l_j}}$
   
\caption{Cross-validation Variable Selection in Component-wise Gradient Boosting}
\label{algo: cv boost}
\end{algorithmic}
\end{algorithm}

\bigskip


\subsection{Akaike Information Criterion (AIC)}
\label{sec: AICboost}

As a second option the AIC can be used in the variable selection step. This approach has been successfully integrated in likelihood-based boosting for Generalized Linear Mixed Models \cite{tutz2010generalized}.
In order to include the {AIC} into the boosting procedure, a definition of the degrees of freedom (df) is needed. In a usual multivariate {GLM} the {df} correspond to the number of included parameters $k+1$ which equals the rank of the design matrix $\boldsymbol X$. The common way to derive the {df} of a {GLM} is to compute the trace of the hat matrix.
Thus, a specification of the hat matrix for the boosting algorithm is needed which has been derived by B\"uhlmann and Hothorn \cite{Buhlmann.2007} and is displayed in Equation \eqref{eq: B_t long}. We restrict the explanation to the {df} for {OLS} base learners, since other base learners are not applied throughout the paper.
\begin{eqnarray}
\label{eq: B_t long}
\boldsymbol B^{(t)} &=& \boldsymbol B^{(t-1)} + \nu \cdot \boldsymbol H_{j^*_t} \left(\boldsymbol I - \boldsymbol B^{(t-1)}\right) \nonumber\\
&=& \boldsymbol I - \left (\boldsymbol I - \nu \boldsymbol H_{j^*_t}\right) \cdot  \left(\boldsymbol I - \nu \boldsymbol H_{j^*_{t-1}}\right) \cdots \Big(\boldsymbol I - \nu \boldsymbol H_{j^*_{1}}\Big),
\end{eqnarray}
with $j^*_t \in \{1, \ldots, k\}$ representing the selected covariate in iteration $t$. Similarly, $\boldsymbol H_{j^*_t}$ corresponds to the hat matrix of the covariate $j^*_t$ together with an intercept.
To account for the response function $h(\boldsymbol \eta_i)$ used in {GLMs}, a weighting matrix $\boldsymbol W=\text{diag}(w_1,\ldots, w_N)$ with $w_i = (\partial h(\eta_i)) ^2 / \text{var}(y_i)$ has to be included in Equation \eqref{eq: B_t long} resulting in 

\begin{eqnarray}
\label{eq: B_t}
    \boldsymbol B^{(t)} &=& \boldsymbol B^{(t-1)} + \nu \cdot \boldsymbol W \boldsymbol H_{j^*_t } \left(\boldsymbol I - \boldsymbol B^{(t-1)}\right) \nonumber \\
    \text{df}^{(t)} &\approx& \text{tr}(\boldsymbol B^{(t)}).
\end{eqnarray}

The derivation of $\boldsymbol W$ follows from the Maximum-Likelihood estimation of a {GLM} \cite{Fahrmeir2021glm}.
B\"uhlmann and Hothorn emphasized that $\boldsymbol B^{(t)}$ can only be seen as an approximation of the hat matrix, since it depends on $\boldsymbol y$ due to the iterative selection of $j^*$ \cite{Buhlmann.2007}. A comment of Hastie \cite{Hastie2007} points out, that this procedure underestimates the true degrees of freedom still leading to a tendency of {AIC} stopping criteria to overfit the training data.

\begin{algorithm}[ht!]
\begin{algorithmic}
  \For{j in $1:k$}

    \State update $j$-th entry of ${\boldsymbol {\hat \beta}^{(t-1)}}$ by the base learner of the $j$-th variable using step length parameter $\nu$ resulting in ${\boldsymbol {\tilde \beta}^{(t)}_j}$
    
    \vspace{9pt}
    \setlength{\parindent}{50pt}
    base learner: ${\hat \beta_j^{(t)}}= \left(\left(\boldsymbol{x}_{(j)}\right )'\boldsymbol{x}_{(j)}\right)^{-1} \left(\boldsymbol{x}_{(j)} \right)' \boldsymbol{u}^{(t)}$
    \vspace{9pt}

    \State evaluate $\text{AIC}_j= -2 \times \text{log} \ \mathcal{L} \left({\boldsymbol {\tilde \beta}^{(t)}_j}\right) + 2 \times \text{df}^{(t)}_{j}$
    
    \EndFor
    
     \vspace{9pt}
    \State choose $\displaystyle{j^* = \underset{j ~\in \{1,\ldots,k\}}{\operatorname{argmin}} ~ \text{AIC}_j}$ and calculate $\boldsymbol B^{(t)}$ from Eq. \eqref{eq: B_t}
 \caption{{AIC} Variable Selection in Component-wise Gradient Boosting}
 \label{algo: AIC boost}
\end{algorithmic}
\end{algorithm}

Despite its imprecise measure of the {df}, the {AIC} still penalizes more complex models stronger and is hence used as an extension of approaches solely based on the loss function. Therefore, the {AIC} is taken as a second option of the variable selection criterion and can be implemented as seen in Algorithm \ref{algo: AIC boost}. The degrees of freedom $\text{df}^{(t)}_{j}$ can be calculated as shown in Equation \eqref{eq: df_kt}.

\begin{eqnarray}
\label{eq: df_kt}
    \boldsymbol B^{(t)}_{j} &=& \boldsymbol B^{(t-1)} + \nu \cdot \boldsymbol W \boldsymbol H_{j} \left(\boldsymbol I - \boldsymbol B^{(t-1)}\right) \nonumber \\
    \text{df}^{(t)}_{j} &\approx& \text{tr}\left(\boldsymbol B^{(t)}_{j}\right).
\end{eqnarray}

For the purpose of shorter computation time, we may apply
$\text{tr}(\boldsymbol A \boldsymbol B) = \text{sum}(\boldsymbol A \cdot \boldsymbol B)$ in the inner $j$-loop with $(\cdot)$ referring to element-wise multiplication because these calculations only require the \emph{trace} of $\left (\boldsymbol H_{j} \left(\boldsymbol I - \boldsymbol B^{(t-1)}\right) \right)$. Nevertheless, as $\boldsymbol B^{(t-1)}$ is needed for calculations in $t$ (and not only its trace), one matrix multiplication of the $N \times N$ matrices shown in Equation \eqref{eq: B_t} has to be performed in each iteration $t$.

Replacing the {AIC} in Algorithm \ref{algo: AIC boost} with the loss function would yield the classical model-based component-wise gradient boosting algorithm.

\section{Simulations}
\subsection{Setup}
\label{sec: data base}

The performance of the two new prediction-based variable selection mechanisms is examined in a simulation study. The benchmark model is generated via \texttt{mboost} \cite{mboost2} using the \texttt{glmboost} command with 10-fold {CV}. The study investigates various scenarios with fixed number of observations $N = 250$, where the values of the design matrix are drawn from a multivariate normal distribution with different settings, i.e.\ 
$$
\boldsymbol X \sim \text{MVN}( \boldsymbol 0, \boldsymbol \Sigma),
 \text{ where } \boldsymbol \Sigma  \in  \left \{\boldsymbol{I}_k, \boldsymbol{\Sigma}_k^{\text{Toep}}  \right \} \text{ and } k \in \{100, 500\}.
$$
$\boldsymbol{I}_k$ represents the identity matrix and $\boldsymbol{\Sigma}_k^{\text{Toep}}$ holds the elements $\text{Cov}\left(\boldsymbol{x}_{(p)}, \boldsymbol{x}_{(q)} \right)=0.9^{|p-q|}, \forall~ p\in \{1,\ldots,k\}$ and $q\in \{1,\ldots,k\}$.
An intercept column is added to $\boldsymbol X$.
Further, the outcome variable is set to be normally distributed
$$y_i = \beta_0 x_{i0} + \sum_{r=1}^{\text{INF}} x_{ir} \beta_r + \varepsilon_i \text{ with } \varepsilon_i \overset{\text{iid}}{\sim} \mathcal N(0,1) \text{ and } \beta_0=1$$
and $\text{INF} \in \{5,20\}$.
The true coefficients are scaled, in order to achieve pre-defined levels of the noise-to-signal ratio (NSR) \cite{Hepp.2016}. The scaling works as follows

$$\beta_r = \kappa \tilde\beta_r , \text{ where } \tilde \beta_r \sim \mathcal{U}(\{-3, -2, -1, 1, 2, 3\}) $$

with the scaling factor $\kappa$, which is defined as $$ \kappa = \sqrt{\frac{1}{\text{NSR} \boldsymbol{\tilde\beta} \boldsymbol \Sigma \boldsymbol{\tilde\beta}}} \text{, where } \boldsymbol {\tilde\beta} = \big ( \tilde \beta_r \big )_{1 \leq r \leq \text{INF}} \text { and } \text{NSR} =\frac{\mathrm{Var}(\boldsymbol\varepsilon)}{ \mathrm{Var}(\boldsymbol{X\beta})} \in \{0.2, 0.5, 1\}.$$

Thus, the simulation study uses a varying number of covariates $k \in \{100,500\}$, varying {NSRs} $\text{NSR} \in \{0.2, 0.5, 1\}$, different amounts of true positive covariates ${\text{INF}} \in \{5, 20\}$ and two correlation structures of the covariates $\boldsymbol \Sigma \in \left \{\boldsymbol{I}_k, \boldsymbol{\Sigma}_k^{\text{Toep}}\right \}$.
The configurations result in 12 low-dimensional ($k < N$) and 12 high-dimensional ($k >N$) settings, as seen in Table \ref{tab: FPRTPR}.

The number of true positives (TPs) covariates ${\text{INF}}$ is chosen to vary, in order to provide different levels of true sparsity. To simulate settings in which the true influential covariates vary in difficulty of identification, the {NSR} is selected to vary from low (easier detection) to high (harder detection). Further, two types of correlation structure are examined to uncover possible different behaviours of the algorithms.
For every setting 100 simulation runs are executed.
To evaluate the out-of-sample predictive performance the models are tested on 100 additional data points evolving the same data generating process. The chosen evaluation criteria regarding variable selection properties and quality of prediction are the {FPR} and the mean squared prediction error (MSPE), respectively. Further the true positive rate (TPR), the number of iterations and the ratios of the average computation time are evaluated.

All algorithms are trained for $T=3000$ boosting iterations with a step-length parameter $\nu = 0.1$. The baseline model and the newly proposed {CV} variable selection method use ten folds. All models start with an offset of $\bar y$, i.e.\ an intercept-only model. 

The following section evaluates the performance of the two modifications in the simulation study and compares them to the benchmark. To emphasize that \texttt{mboost} employs the prediction criteria after fitting the model, whereas the modifications directly include their calculation during the fitting process, the criteria are added at the end ($\text{mboost}_{\text{CV}}$) or at the beginning (AIC-boost, CV-boost) of their names, respectively.


\subsection{Results}


\label{sec: eval}

The results of the simulations are split up into the examined indicators and are displayed and described in the following. The included figures show results for the uncorrelated cases and plots covering both correlation structures can be found in the Appendix \ref{append sim_norm}.

\subparagraph{False positive rate}

\begin{figure}[ht!]
    \centering
    \includegraphics{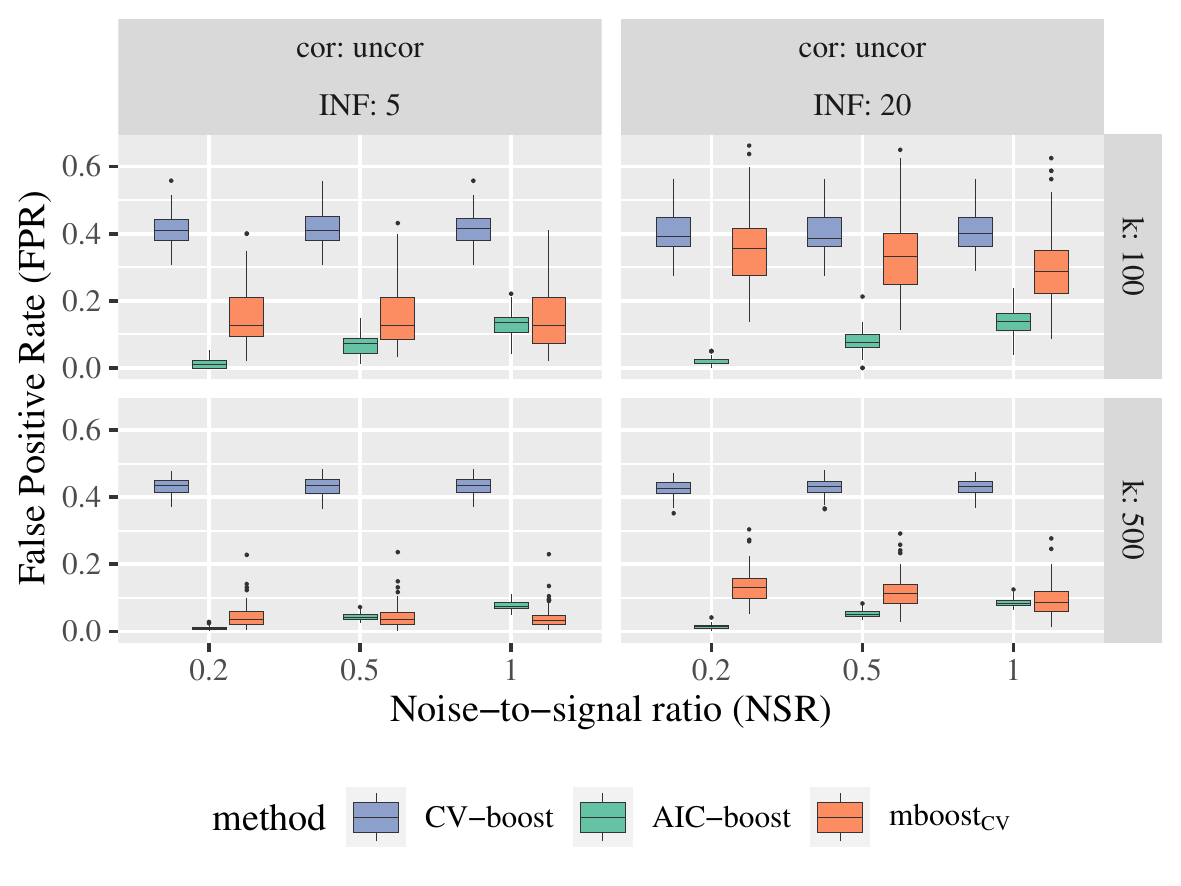}
    \caption{False Positive Rate for new variable selection strategies by simulation setting.}
    \label{fig: FPR color}
\end{figure}

Regarding the {FPR}, we observe the following trend visualized in Figure \ref{fig: FPR color}: The AIC-variable selection (AIC-boost) outperforms both, \texttt{mboost} with 10-fold {CV} ($\text{mboost}_{\text{CV}}$) and CV-boost in most of the tested scenarios, e.g. in case of uncorrelated covariates with $k=500$, ${\text{INF}}=5$ and $\text{NSR}=0.2$ (bottom left, first x-axis entry). In this setting, {CV}-boost results in a median {FPR} of around 0.43, while $\text{mboost}_{\text{CV}}$ yields an {FPR} of 0.04. The median {FPR} of the {AIC}-method is 0.01 in the described setting, which is the lowest of the three algorithms.

This pattern can be found in nearly all tested settings of the simulation study, however little differences are worth to highlight: the variability of the {FPR} of the two new algorithms is substantially lower than $\text{mboost}_{\text{CV}}$ in every setting. Due to the relative nature of the {FPR}, one has to be cautious with comparisons between the low-dimensional and the high-dimensional setting. The variability of the {FPR} in the high-dimensional setting appears smaller, but equal measures of location and scale of the {FPR} would yield different absolute values (i.e.\ {FPs}) depending on the dimensionality.

Another observation concerns the true sparsity of the models, which is varied by $\text{INF}$. It appears, that the performance of $\text{mboost}_{\text{CV}}$ depends on the true sparsity. Enlarging the number of true informative covariates and keeping the other parameters constant results in a higher {FPR} of $\text{mboost}_{\text{CV}}$, whereas the two modifications perform more robust and reveal very similar {FPR} across the differing numbers of informative covariates. This pattern is more pronounced in the uncorrelated settings (see column 1 and 2 in Figure \ref{fig: FPR full}).
\begin{center}
\begin{table*}[ht!]%
\caption{Comparison of False Positive Rates of AIC-boost and $\text{mboost}_{\text{CV}}$ by noise-to-signal ratio.\label{tab: fpr}}
\centering
\begin{tabular*}{500pt}{@{\extracolsep\fill}lcc@{\extracolsep\fill}}
\toprule
&\multicolumn{2}{@{}c@{}}{\textbf{settings with superior performance$^*$ of AIC-boost compared to $\text{mboost}_{\text{CV}}$}} \\\cmidrule{2-3}
  & \textbf{Number}  & \textbf{Percentage}   \\
\midrule
NSR = 0.2 & 8/8    & 100\% \\ 
NSR = 0.5 & 7/8    & 88\%  \\
NSR = 1 & 4/8    & 50\%  \\ 
\bottomrule
\end{tabular*}
\item[$^*$] Superior performance is measured by means of the median {FPR}.
\end{table*}
\end{center}
The performance of the {AIC}-method looks promising, since it outperforms $\text{mboost}_{\text{CV}}$ in 19/24 ($79\%$) settings in terms of the median {FPR}. A detailed listing can be found in Table \ref{tab: fpr_full}. It summarizes the number of settings by parameter in which AIC-boost has a lower median {FPR} than the benchmark $\text{mboost}_{\text{CV}}$. The most influential parameter appears to be the {NSR} (see Table \ref{tab: fpr} and Table  \ref{tab: fpr_full}). {AIC}-boost always results in sparser models when $\text{NSR}=0.2$ is used, regardless of high or low dimension and the amount of informative covariates. Contrary, simulation settings with $\text{NSR}=1$ result in a share of only 4/8 ($50\%$) settings, where {AIC}-boost outperforms the benchmark model. Thus, in situation with strong signal, AIC-boost is more likely to outperform $\text{mboost}_{\text{CV}}$ than in weak signal situations. Generally, AIC-boost performs better in settings with a high signal compared to weak signal settings. A reason for this may be the difference between the contributions of {TPs} and {FPs}: In strong signal situations, the contribution of the true positives to minimize the loss function exceeds the smaller contributions of FP to a large extend. The penalty term of the {AIC} ensures, that the model refrains from including them, until the contribution of the {TP} becomes small enough. In weak signal situations the contributions of {TPs} and {FPs} do not differ so much and thus, the algorithm includes more {FPs}.

In terms of the {FPR}, the simulation study clearly recommends using the newly tested {AIC}-boost in low-dimensional settings for both correlation structures. In high-dimensional data settings, the performance is determined by the {NSR}. However, in 8/12 high-dimensional settings {AIC}-boost outperforms $\text{mboost}_{\text{CV}}$. The variability of the {FPR} is drastically reduced in every tested simulation set-up (see Figure \ref{fig: FPR color} and Table \ref{tab: FPRTPR}), which suggests that {AIC}-boost is a more robust estimation procedure in terms of the {FPR}.

\begin{table*}[ht!]%
\caption{Median (variance $\times$ 100) FPR and TPR for AIC-boost and $\text{mboost}_{\text{CV}}$ by simulation setting.\label{tab: FPRTPR}}
\centering
\begin{tabular*}{500pt}{@{\extracolsep\fill}c|l|c|c|cccc@{\extracolsep\fill}}
\toprule
&&&&\multicolumn{2}{@{}c@{}}{\textbf{FPR}} & \multicolumn{2}{@{}c@{}}{\textbf{TPR}} \\\cmidrule{5-6}\cmidrule{7-8}
\textbf{correlation} & \textbf{k}  & \textbf{INF} & \textbf{NSR} & \textbf{AIC-boost} & \textbf{$\text{mboost}_{\text{CV}}$} &  \textbf{AIC-boost} & \textbf{$\text{mboost}_{\text{CV}}$}   \\
\midrule

\multirow{12}{*}{uncorrelated} & \multirow{6}{*}{100} & \multirow{3}{*}{5}& 0.2 & 
\textbf{0.011 (0.014)} &0.126 (0.739) &1.000 (0.000) &1.000 (0.000) \\
& & & 0.5 & 
\textbf{0.074 (0.085)} &0.126 (0.737) &1.000 (0.000) &1.000 (0.000) \\
& & & 1 &
0.137 (0.114) &\textbf{0.126 (0.789)} &1.000 (0.055) &1.000 (0.133) \\
\cmidrule(lr){3-8}

&& \multirow{3}{*}{20}& 0.2&
\textbf{0.025 (0.022)} &0.356 (1.215) &0.952 (0.272) &\textbf{1.000 (0.013)} \\
& & & 0.5&
\textbf{0.075 (0.094)} &0.331 (1.263) &0.905 (0.415) &\textbf{0.952 (0.182)} \\
& & & 1 &
\textbf{0.138 (0.162)} &0.288 (1.234) &0.810 (0.420) &\textbf{0.905 (0.483)} \\
\cmidrule(lr){2-8} 

& \multirow{6}{*}{500} & \multirow{3}{*}{5} & 0.2 &
\textbf{0.010 (0.003)} &0.037 (0.120) &1.000 (0.000) &1.000 (0.000) \\
& & & 0.5 &
0.041 (0.010) &\textbf{0.037 (0.124)} &1.000 (0.028) &1.000 (0.082) \\
& & & 1 &
0.075 (0.015) &\textbf{0.032 (0.106)} &1.000 (0.296) &1.000 (0.609) \\
\cmidrule(lr){3-8}

& & \multirow{3}{*}{20} & 0.2&
\textbf{0.015 (0.004)} &0.131 (0.210) &0.952 (0.236) &\textbf{1.000 (0.116)} \\
& & & 0.5 & 
\textbf{0.052 (0.011)} &0.115 (0.207) &0.857 (0.424) &\textbf{0.905 (0.529)} \\
& & & 1 &
\textbf{0.085 (0.016) }&0.088 (0.202) &0.762 (0.550) &0.762 (1.077) \\
\midrule

\multicolumn{1}{l|}{\multirow{12}{*}{Toeplitz correlation}} &
\multicolumn{1}{c|}{\multirow{6}{*}{100}} & \multirow{3}{*}{5} &
0.2&
\textbf{0.011 (0.011)} &0.200 (1.418) &1.000 (1.485) &1.000 (1.015) \\

\multicolumn{1}{l|}{} & \multicolumn{1}{c|}{} & & 
0.5 &
\textbf{0.032 (0.039)} &0.158 (1.549) &0.833 (1.588) &0.833 (1.958) \\

\multicolumn{1}{l|}{} &
\multicolumn{1}{c|}{} & & 
1 &
\textbf{0.053 (0.063)} &0.100 (1.288) &0.833 (2.153) &0.833 (2.706) \\

\cmidrule(lr){3-8}
\multicolumn{1}{l|}{} & \multicolumn{1}{c|}{} & \multirow{3}{*}{20} & 
0.2& 
\textbf{0.013 (0.019)} &0.350 (2.203) &0.571 (1.594) &\textbf{0.833 (2.109)} \\

\multicolumn{1}{l|}{} & \multicolumn{1}{c|}{} & & 
0.5 & 
\textbf{0.038 (0.059)} &0.200 (2.178) &0.571 (1.823) &\textbf{0.667 (2.579)} \\

\multicolumn{1}{l|}{} & \multicolumn{1}{c|}{} & & 
1 & 
\textbf{0.050 (0.090)} &0.125 (1.842) &0.524 (1.420) &0.524 (2.394) \\

\cmidrule(lr){2-8}
\multicolumn{1}{l|}{} & \multirow{6}{*}{500} & \multirow{3}{*}{5} & 
0.2 &
\textbf{0.007 (0.002)} &0.069 (0.368) &\textbf{1.000 (1.031)} &0.833 (1.915) \\

\multicolumn{1}{l|}{} & & & 
0.5 & 
\textbf{0.024 (0.007)} &0.043 (0.230) &\textbf{1.000 (1.459)} &0.833 (2.666) \\

\multicolumn{1}{l|}{} & & & 
1 &
0.042 (0.009) &\textbf{0.020 (0.174)} &\textbf{0.833 (1.948)} &0.667 (2.547) \\

\cmidrule(lr){3-8}
\multicolumn{1}{l|}{} & & \multirow{3}{*}{20} & 
0.2 &
\textbf{0.008 (0.005)} &0.073 (0.476) &0.571 (1.814) &0.571 (1.630) \\

\multicolumn{1}{l|}{} & & & 
0.5 &
\textbf{0.025 (0.009)} &0.035 (0.185) &\textbf{0.524 (1.452)} &0.429 (1.238) \\
\multicolumn{1}{l|}{} & & & 
1 &
0.044 (0.016) &\textbf{0.020 (0.131) }&\textbf{0.476 (1.197)} &0.381 (1.290) \\
\bottomrule
\end{tabular*}
\end{table*}

\newpage
\subparagraph{True positive rate}

Another trend can be observed when looking at the {TPR} in Table \ref{tab: FPRTPR} and Figure \ref{fig: TPR}: In settings with just ${\text{INF}}=5$ informative covariates, the {AIC}-boost yields very good (median) {TPR}, which are always better or equal compared to $\text{mboost}_{\text{CV}}$. In settings with ${\text{INF}}=20$ informative covariates {AIC}-boost reveals the tendency to have a lower median {TPR} than the baseline method (7/12 scenarios). Differences are less pronounced in the uncorrelated settings ($\approx$ five percentage points worse). Note, the {TPR} can be compared between low- and high-dimensional settings, as it only depends on INF and not on $k$. Using 20 informative Toeplitz-correlated covariates, {AIC}-boost outperforms $\text{mboost}_{\text{CV}}$ in the high-dimensional setting, whereas this relationship is reversed in the low-dimensional cases.

{CV}-boost does not show a clear trend in terms of {TPR}. It eventually outperforms {AIC}-boost and the benchmark model, e.g. Toeplitz correlated case with $k= 500, \text{NSR}=0.2, {\text{INF}}=20$ and sometimes reveals a {TPR} similar to {AIC}-boost. While the {FPR} is clearly reduced by {AIC}-boost, this is not recognizable with the same degree of certainty for the {TPR}.

In summary, the much lower {FPR} for {AIC}-boost is accompanied by a slightly lower {TPR} in some setups.

\subparagraph{Mean squared prediction error}

\begin{figure}[ht!]
    \centering
    \includegraphics{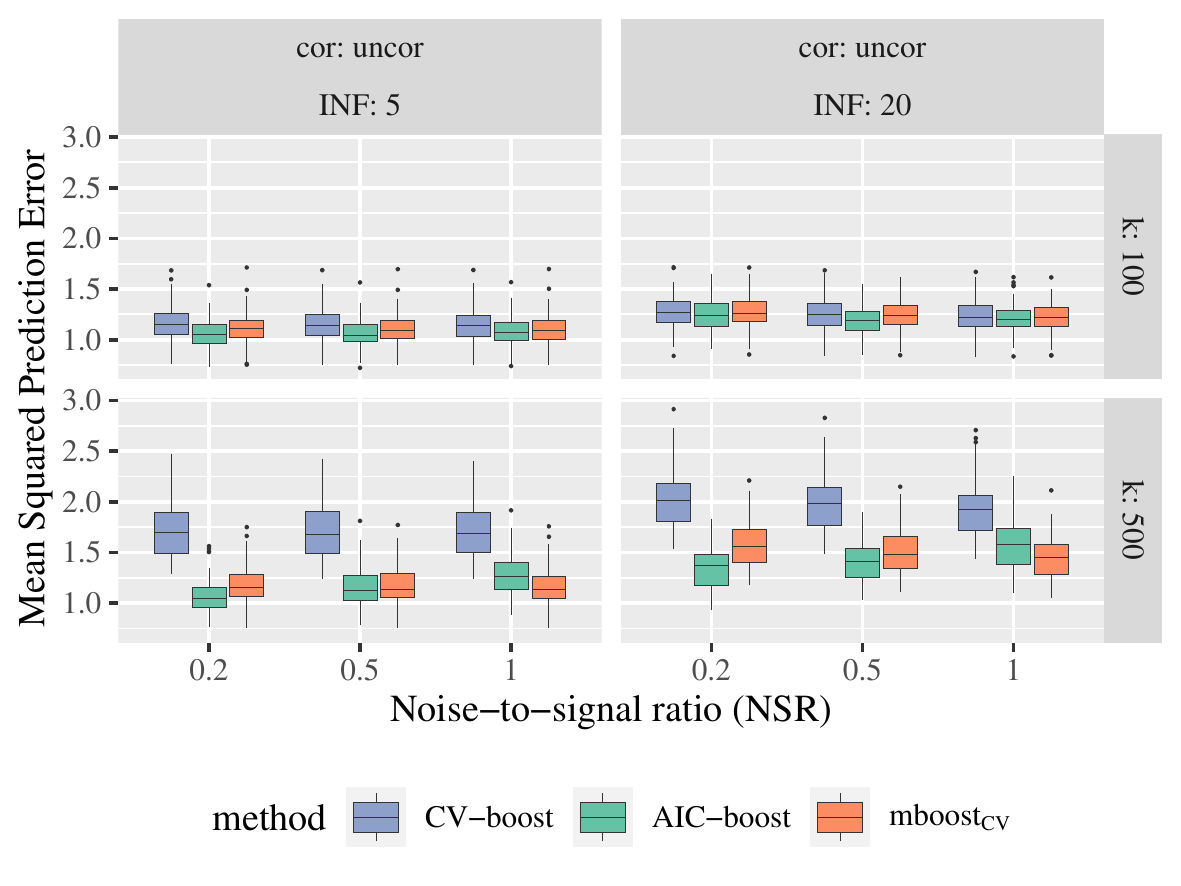}
    \caption{Mean Squared Prediction Error for new variable selection strategies by simulation setting.}
    \label{fig: MSPE}
\end{figure}

Since prediction and sparsity were seen as two opposing aims in model selection processes, the next indicator to analyze is the {MSPE}. Despite the fact that {AIC}-boost often results in sparser models, which increases the interpretability, it still keeps up with the benchmark model in terms of prediction accuracy (see Figure \ref{fig: MSPE}). CV-boost produces slightly (considerably) higher median {MSPE} in low-dimensional (high-dimensional) settings with uncorrelated covariates. Using the Toeplitz covariance structure, differences between the three methods are less pronounced. Thus, applying the {AIC}-boost approach results in sparser models without sacrificing the predictive performance of the model in the vast majority of scenarios.

\begin{figure}[ht!]
    \centerline{\includegraphics{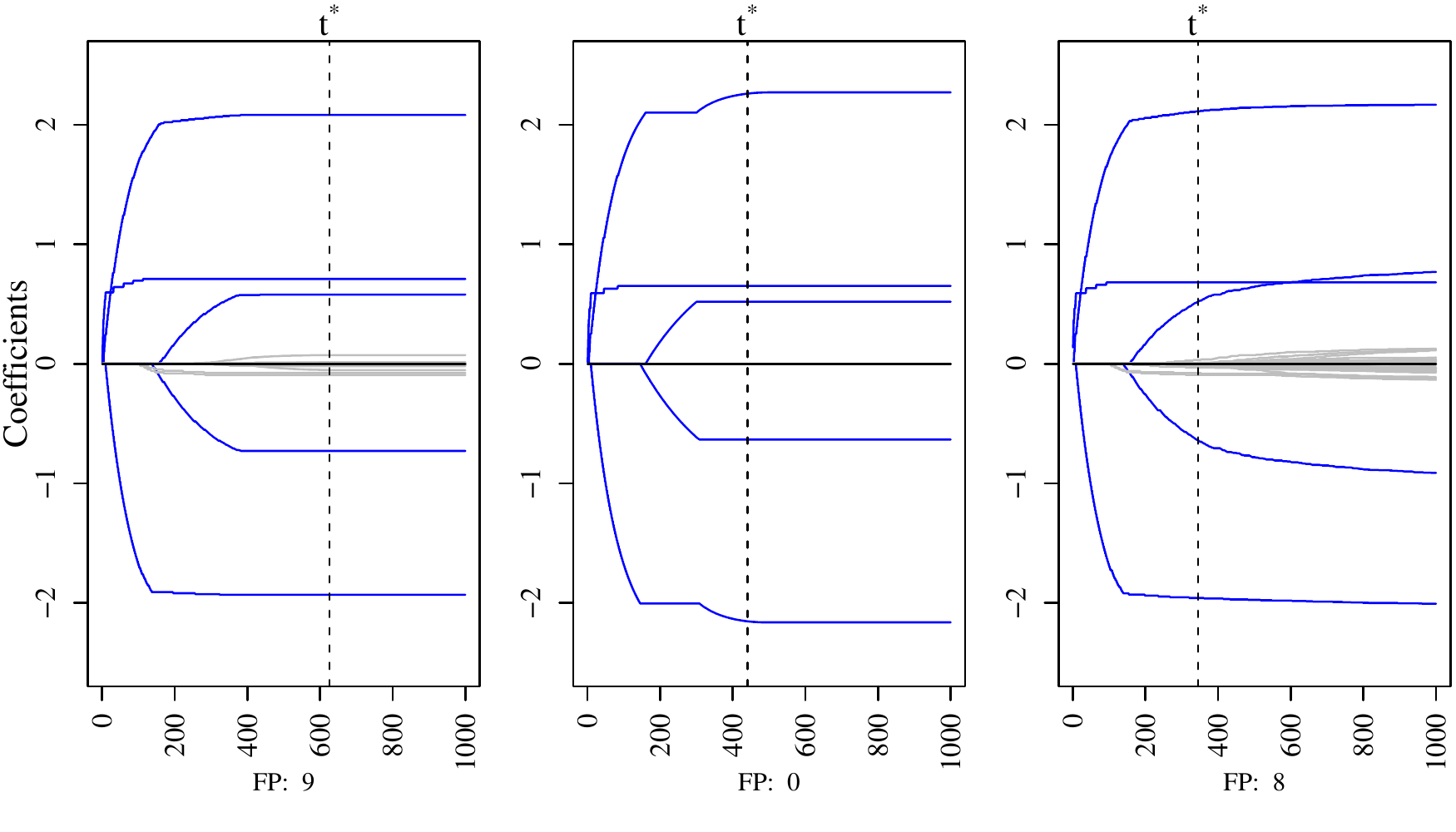}}
    
    \caption{Coefficient paths of CV-boost (left), AIC-boost (middle) and $\text{mboost}_{\text{CV}}$ (right). Blue paths correspond to true influential covariates, grey represents false positives.} 
    
    \label{fig: pathplots}
\end{figure}

\subparagraph{Number of iterations and computation time}
Looking at the selected stopping iteration (Figure \ref{fig: SI}), $\text{mboost}_{\text{CV}}$ always exhibits the lowest median in terms of stopping iteration. After looking at the higher {FPR} of $\text{mboost}_{\text{CV}}$ one would probably expect the contrary -- a later stopping iteration, meaning more possibilities to include {FPs}. But since the modification of the variable selection step changes the pattern in the coefficient paths (see Figure \ref{fig: pathplots}), earlier stopping iterations of $\text{mboost}_{\text{CV}}$ can yield more included covariates when compared to {AIC}-boost at the same iteration.
Figure \ref{fig: pathplots} illustrates coefficient paths of the two modifications and the baseline model exemplarily. The figure results from one simulation run with Toeplitz correlated covariates in a Gaussian setting using $k=100$, $\text{INF}=5$ and $\text{NSR}=0.2$. It can be seen, that changing the variable selection step influences how many covariates enter the final model. Further, the size of the coefficients and the order of inclusion may be affected. The coefficient paths of AIC-boost (Figure \ref{fig: pathplots} middle) are the steepest and its resulting final model includes the least {FPs}.


The most influential parameter of the simulation governing the computation time of the algorithms (beside the number of observations which is held constant) is the number of possible covariates $k$. Table \ref{tab: comptime} displays the ratio of the average computation time for the two new algorithms compared to $\text{mboost}_{\text{CV}}$ in low and high-dimensional settings. The algorithms were run on an AMD EPYC 7742 Processor with two sockets. Each socket includes 64 cores with two threads per core and overall 512 GB RAM.
\begin{center}
\begin{table}[ht!]%
\centering
\caption{Ratio of average computation time of one simulation run by number of covariates (algorithm/$\text{mboost}_{\text{CV}}$).\label{tab: comptime}}%
\begin{tabular*}{500pt}{@{\extracolsep\fill}lcc@{\extracolsep\fill}}
\toprule
\textbf{} & \textbf{CV-boost}  & \textbf{AIC-boost} \\
\midrule
k=100 &  1151  & 142 \\
k=500 & 9261 & 310 \\
\bottomrule
\end{tabular*}
\end{table}
\end{center}

Clearly, both new algorithms take longer than the benchmark method. As expected CV-boost takes the longest.
Note, that besides the calculation rules for {AIC}-boost mentioned above, no attempts were made to make the algorithms more efficient. Contrary, \texttt{mboost} is an optimized package.

\section{Data}
\subsection{Data Explanation}

\begin{figure}[ht!]
    \centerline{\includegraphics{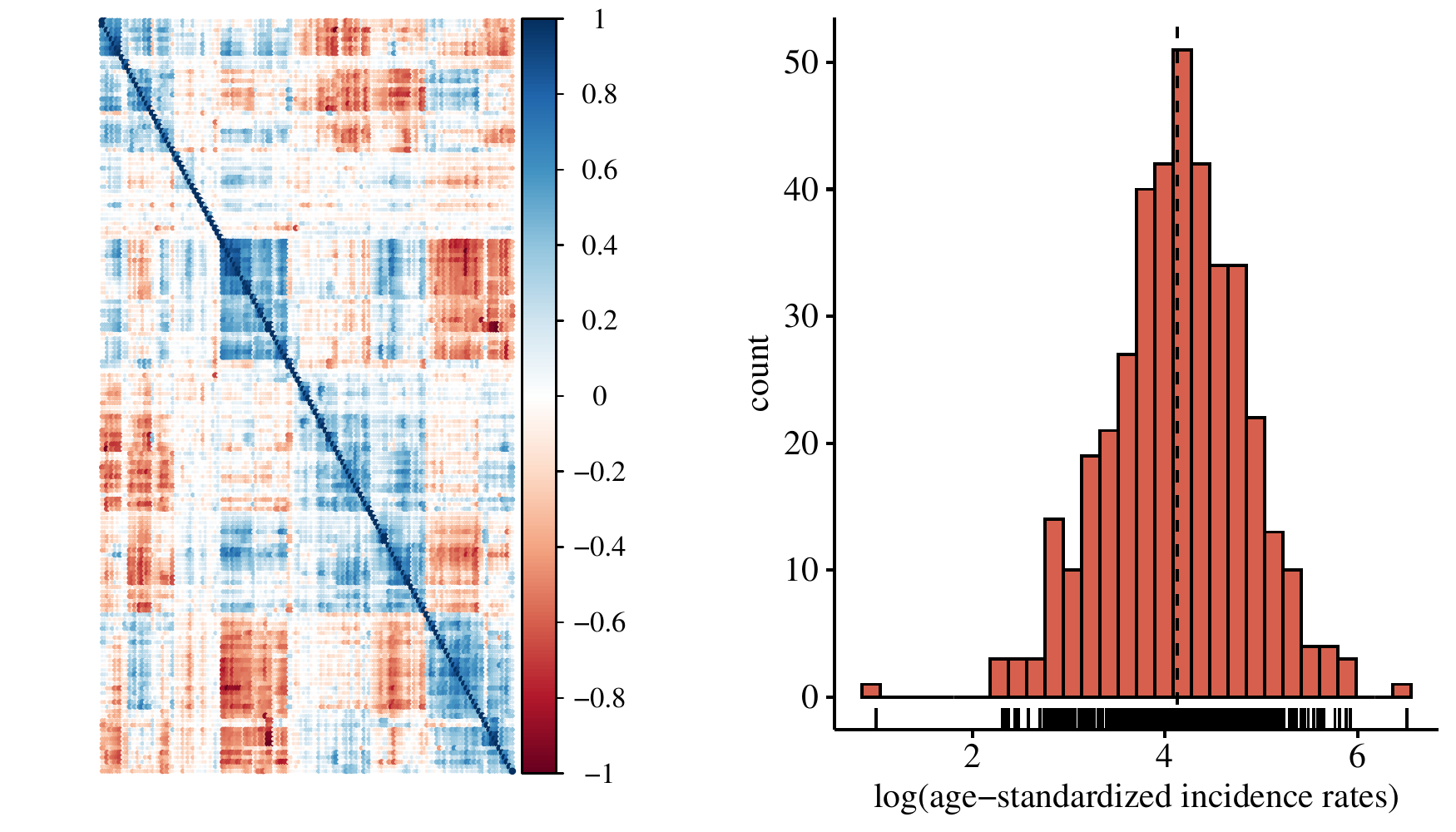}}
    \caption{COVID-19 data set descriptives: correlation matrix of COVID-19 data set (left), empirical distribution of log(age-standardized incidence rates) (right).} 
    \label{fig: covid2}
\end{figure}

To evaluate the two modified algorithms in a more realistic setting, they were applied to a real world data set.
Therefore, a subset of a COVID-19 data base from Doblhammer et al. is used \cite{Doblhammer2022}. The authors investigated the relationship of county-scale variables on the county-specific age-standardized COVID-19 incidence rates in Germany in order to uncover possible social disparities. In this framework, they focussed on different periods of the pandemic in Germany and identified the ten most important county-scale covariates for each of the five periods of the pandemic in Germany by means of {SHAP} values. The used subset of their data contains 163 variables measured on 401 counties in Germany. The variables cover different socioeconomic characteristics, like demography, social economic and settlement structure, health care, poverty, unemployment, politics and education, interrelationship with other regions, e.g. percentage change of persons from 2012 to 2017, old-age (65+) dependency ratio (2017), total fertility rate (2017), unemployment rate (2017), persons per square kilometer (2017) or share of area in natural state (2017).
By looking at the correlation structure of the possible covariates in Figure \ref{fig: covid2} (left), one can observe highly (positive and negative) correlated blocks of variables (e.g. ``share of employed persons in tertiary sector in all dependently employed persons" has a Pearson correlation coefficient of $\approx -0.93$ with ``share of employed persons in secondary sector in all dependently employed persons" and ``unemployment rate" correlates by $\approx 0.94$ with ``share of persons with basic social security benefits per 1,000 persons") corresponding to thematically-related covariates. All variables are either metric or dummy-coded. Metric covariates are standardized.
The outcome of interest is the age-standardized incidence rate on the county-level for the first lockdown period (March 16,2020 - March 31,2020) as defined by Doblhammer et al. \cite{Doblhammer2022}. It follows a log-normal distribution (Figure \ref{fig: covid2} right).

\subsection{The Model}

In order to extract the most influential macro-variables associated with the log-transformed age-standardized COVID-19 incidence rates out of the pool of 163 covariates, we apply the two modified algorithms and compare their performance to \texttt{mboost} with 10-fold {CV} ($\text{mboost}_{\text{CV}}$). For reasons of comparability, $\text{mboost}_{\text{CV}}$  and CV-boost are trained using the same folds. Since these two algorithms were highly dependent on the data split, they are run five times with different data splits and median values are reported.  The calculation of the {MSPE} is based on 100 randomly chosen data points, such that the models are trained with 301 observations.
Note, that the chosen covariates may differ from the ones identified by the original article \cite{Doblhammer2022}, since they used gradient boosting for prediction with tree-based weak learners, whereas this paper applies model-based component-wise gradient boosting with {OLS} base learners. The latter mentioned implies the assumption of a true linear relationship between the covariates and the log-transformed age-standardized incidence rate.

The results contradict the main message of the simulation study in terms of sparsity, as one would expect $\text{mboost}_{\text{CV}}$ to include more variables than {AIC}-boost.

\begin{center}
\begin{table}[ht!]%
\centering
\caption{Performance of algorithms on COVID-19 data set.\label{tab: covid_res}}%
\begin{tabular*}{500pt}{@{\extracolsep\fill}lcccc@{\extracolsep\fill}}
\toprule
\textbf{method} & \textbf{stopping iteration}  & \textbf{no. of covariates}  & \textbf{MSE}  & \textbf{MSPE} \\
\midrule
$\text{mboost}_{\text{CV}}$ $\dagger$  & 20    & 8 & 0.232 & 0.365  \\
$\text{mboost}_{\text{AIC}}$ & 2039  & 85 & 2.553  & 2.801   \\
LASSO-AIC  & /     & 10  & 0.211  & 0.342   \\
&&&& \\
AIC-boost        & 164   & 21  & 0.154  & 0.296   \\
CV-boost $\dagger$         & 447   & 53   & 0.140  & 0.281   \\
\bottomrule
\end{tabular*}
\item[$\dagger$] model averaging performed, median values are reported.
\end{table}
\end{center}

However, $\text{mboost}_{\text{CV}}$ includes only nine covariates while the model using {AIC}-boost contains 21 covariates. This high sparsity of $\text{mboost}_{\text{CV}}$ comes along with a poorer predictive performance regarding the {MSE} and {MSPE}. Both tested algorithms outperform $\text{mboost}_{\text{CV}}$ regarding the predictive performance but include (many) more variables than the baseline model. They also outperform {LASSO} with an {AIC} stopping criterion in terms of prediction accuracy. Comparing only the two newly tested algorithms, the predictive performance is very similar by differing numbers of included covariates. Applying Occam's razor, AIC-boost combines the sparsest model with a comparable prediction accuracy and thus would be the preferred model in this application.

\subsection{Interpretation of Coefficients}

During the first COVID-19 wave, there was a change in the association of socioeconomic status (SES) and incidence rates. A positive correlation could be shown between SES and high incidence rates in the early phase, whereas higher incidence rates were associated with low SES in the later phase \cite{Plumper.2020, Wachtler.2020}. Reasons for this included higher mobility (e.g. skiing vacations) in the higher SES groups in the early periods before the introduction of government mitigation measures and less flexible working conditions in the lower SES groups later in the first wave \cite{Rohleder.2022}. The period from March 16 to March 31 was marked by strict lockdown measures to avoid personal contact.
Table \ref{tab: coef covid} displays the covariates and their coeﬀicients for the three sparsest models. For $\text{mboost}_{\text{CV}}$ five of the 8 selected variables are consistent with the ten most important variables identified in the original article \cite{Doblhammer2022} and share the same direction of influence (positive/negative) on the outcome variable with the {SHAP} value-based approach. Two further selected variables (``Persons in long-term care per 10.000 persons in 2017” and ``\%Households with low income (1,500€ per month) in all households in 2016”) are equivalent indicators (Care\_allowance\_receivers and Social\_security\_benefits) covering same dimensions (poor health and households with low income). Only the variable ``Premature mortality” is not covered by Doblhammer et al. \cite{Doblhammer2022}, whereby premature mortality is higher in socioeconomically weaker areas \cite{Plumper.2018} and is thus consistent with the background of a positive social gradient.
The ranking of covariates in terms of the absolute value of the coeﬀicients is very similar between {LASSO} and $\text{mboost}_{\text{CV}}$. However, the {LASSO} selection contains two more variables. The variable ``Ever 100+ inbound commuters from Tirschenreuth” represents the connection/mobility to county of Tirschenreuth, which was a hotspot area with the highest incidences at this period, and increased even further since the start of March \cite{Brandl.2020}. The variable ``\%Older employed persons (55 years+) in all employed persons” represents aging differences between the counties which is accompanied by structural weakness. This may indicate a positive social gradient as well as a geographical aspect. In recent years, aging and population shrinking in the rural eastern regions has been faster than in other rural areas \cite{Fuest.2019}; whereas, incidences were lower there than in other regions during the observation period.
AIC-boost incorporates several coefficients with a rather large absolute value, that are not even included in the other model. Three of these variables regard outbound commuters to hotspot areas during this period (``Ever 100+ inbound commuters\dots"). The variables concerning the number/share of employed persons in the younger/older/general population (``\%Young employed persons in all young persons (under 26 years) in 2017”, ``\%Older employed persons in all older persons (55 years+) in 2011-2017”, and ``\%Change of number of employed persons in 2012-2016”) indicate the level of employment and are consistent with a positive social gradient, indicated by the positive sign. While there tended to be no differences in incidence between the sexes during the course of the pandemic \cite{Ballering.2021}, differences in infection rates between men and women appeared in the early phase of the pandemic \cite{Bianconi.2020}. Younger women had higher infection rates during this phase \cite{Doerre.2022, Ancochea.2021}. The variable ``\%Change of number of persons at age 50-65 in 2012-2017” indicates population growth which is stronger in the south respectively partly negative in rural areas in eastern Germany where structural and economic indicators are low \cite{Leibert.2022, Fink.2019}. The variables ``Average travel time to the next large-sized regional center (Oberzentrum)” and ``\%Outbound commuters…” represent the rurality of counties. In contrast to the orginal article \cite{Doblhammer2022}, indicators for rurality are correlated with lower incidence rates here.

Note, that no causal relationships on a sociological micro-level can be extracted here, since both, the outcome variable and the covariates are macro-indicators. Ignoring this, would lead to the ecological fallacy \cite{Robinson1950}.

\section{Conclusion}

Based on promising findings \cite{tutz2010generalized, buhlmann2006sparse} and due to the lack of a systematic investigation on different variable selection mechanisms in gradient boosted {GLMs}, this paper evaluated two modifications of component-wise gradient boosting. 
Instead of minimizing the loss function in model-based boosting, two prediction-based criteria were used instead, namely: cross-validation (CV-boost) and Akaikes Information Criterion (AIC-boost). The new modifications directly minimize these criteria and were expected to reduce the false positive rate (FPR). 

The simulation study with a normally distributed outcome revealed that {CV}-boost does not outperform the benchmark model $\text{mboost}_{\text{CV}}$ in terms of the {FPR}. AIC-boost yields promising results, outperforming the benchmark {FPR} in 79\% of the simulation settings (measured by median). Simultaneously, it provides comparable results in terms of predictive performance (median {MSPE}). Even though it sometimes decreases the {TPR} to a small extend, its reduction of  median and variance of the {FPR} are indicators for promising results in further research. The most influential parameter governing the performance of AIC-boost appears to be the noise-to-signal ratio (NSR). The lower the signal in relation to the noise, the poorer the performance of AIC-boost. This result is independent of the setup's dimensionality, correlation structure and number of informative covariates. The largest improvements of the {FPR} were observed in less sparse models (i.e.\ INF $=20$). Further research on the performances in non-sparse situations may contribute to a proper evaluation of AIC-boost. The two modifications have been applied to a $k<N$ real world data set. {AIC}-boost outperforms the benchmark model in this applications in terms of prediction and yields the sparsest model when compared to models with similar predictive performance. The results are plausible and widely consistent with other research in terms of selected variables and sign of coefficients. Although we used a nearly identical subset of data \cite{Doblhammer2022}, especially AIC-boost selected some different variables. However, most of them are only other indicators for a similar or the same dimension identified by the former study. A reason for the differences may the fact that our OLS base learners assumed linear relations while tree-based methods used in the original article also account non-linear associations and possible interactions. 
In summary, the findings suggest that the {AIC} modification can improve variable selection properties in component-wise gradient boosting by bridging sparsity and predictive performance. The purely loss-function based approaches CV-boost does not exhibit a lower {FPR} and further result in worse predictions.

These results, however, have certain limitations. First, the number of observations in the simulation study was kept constant at $N=250$. In order to test the stability of the {AIC}-boost performance, it is interesting to analyze the behaviour with less and more observations.
Second, the modifications are restricted to the {OLS} base learner; however, other base learners, such as splines or tree-based ones, are worth investigating. This is also important in order to preserve the flexibility of the boosting framework. If the presented AIC-boost approach is expanded to other base learners, the algorithm's preference for less flexible base learners in later iterations must be taken into account \cite{Hofner.2011}.

Despite the fact that the simulation study used a common model selection strategy as benchmark, further research comparing {AIC}-boost to other boosting modifications (e.g.\ probing \cite{Thomas.2017} or deselection \cite{stromer2021deselection}) will provide a more complete picture of its performance.
The paper only presented two modifications of the variable selection step, however, other criteria such as bootstrapping, {BIC} or a corrected {AIC} \cite{hurvich1989regression}, can be used and may be interesting to test.

Since the simulation study only addresses one type of outcome-distribution, a simpler simulation study with Poisson distributed values has been performed to overcome this limitation (see Appendix \ref{append sim_pois}). 
Contrary to the expectations, the {FPR} is not reduced in a majority of the settings when applying AIC-boost. A slight tendency of AIC-boost to perform better in less sparse setups, however, is also found using a Poisson distributed outcome. In most settings AIC-boost results in higher prediction errors. With regard to the {TPR}, $\text{mboost}_{\text{CV}}$ does not perform well, especially in the case of a high {NSR}, and it becomes obvious, that AIC-boost often results in a higher {TPR}. In this simulation, the higher sparsity of $\text{mboost}_{\text{CV}}$ comes with the drawback of a low {TPR} which is undesirable. AIC-boost seems to perform a more conservative variable selection here, as the inclusion of more {TPs} comes with a larger model, including more {FPs} as well.
The {NSR} does not govern the performance of AIC-boost to a large extend, as it has been observed for the normal distribution.
Note, that the {NSR} cannot be calculated and induced directly as in the case of a normally distributed outcome. As a replacement, different sizes of possible coefficients were used. However, their absolute level in comparison to the simulation study above remains unclear and only the relative difference between the three levels can be used as a proxy for the {NSR}.

Some first assumptions on the underlying reasons for the differing results of the two simulations include the approximation of the hat matrix, which may be inaccurate in this setup. Since the advantages of using AIC-boost in terms of sparsity and prediction accuracy diminish, the preliminary results warrants further research to test the properties of the modifications for other types of {GLMs}, e.g.\ Binomial distributed outcomes.
Investigations into shorter computation time may be fruitful as well. The use of early stopping in the algorithms might speed them up considerably. In the simulation study this would have reduced the computation time of {AIC}-boost to around 1/6 of the original running time. Furthermore, research into a better approximation of the {df} will be highly appreciated. A good point to start may be the implementation of the cardinality of the active set of covariates as a measure for the {df} \cite{Buehlmann2007rejoinder} and compare the results.

Despite these limitations, this paper serves as a basis for further research on prediction-based variable selection in component-wise gradient boosting.

\section*{Acknowledgments}
We would like to thank Gabriele Doblhammer and Daniel Kreft for aggregating and sharing the data.

\subsection*{Author contributions}

\textbf{Sophie Potts}: Conceptualization; investigation; methodology; formal analysis; software; writing – original draft; writing – review \& editing. \textbf{Elisabeth Bergherr}: Supervision; writing – review \& editing. \textbf{Constantin Reinke}: Validation; writing – review \& editing. \textbf{Colin Griesbach}: Conceptualization; writing – review \& editing.

\newpage

\appendix
\section{Simulation Results \label{app1}}

\subsection{Simulation Results of Normally Distributed Data}
\label{append sim_norm}

\begin{figure}[ht!]
    \centering
    \includegraphics{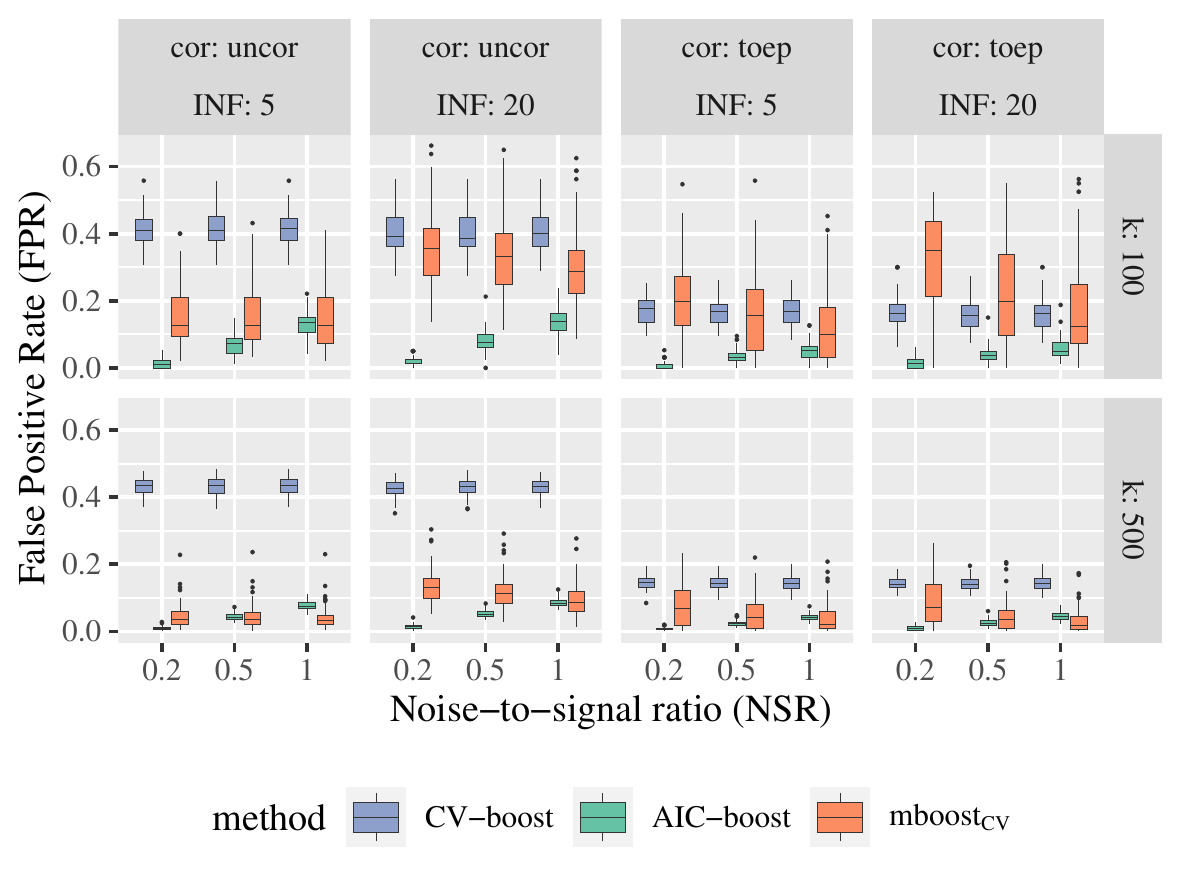}
    \caption{False Positive Rate for new variable selection strategies by simulation setting.}
    \label{fig: FPR full}
\end{figure}

\begin{center}
\begin{table*}[ht!]%
\caption{Comparison of False Positive Rates of AIC-boost and $\text{mboost}_{\text{CV}}$ by simulation parameter.\label{tab: fpr_full}}
\centering
\begin{tabular*}{500pt}{@{\extracolsep\fill}lcc@{\extracolsep\fill}}
\toprule
&\multicolumn{2}{@{}c@{}}{\textbf{settings with superior performance$^*$ of AIC-boost compared to $\text{mboost}_{\text{CV}}$}} \\\cmidrule{2-3}
  & \textbf{Number}  & \textbf{Percentage}   \\
\midrule
uncorrelated            & 9/12   & 75\%  \\ 
Toeplitz correlation    & 10/12  & 83\%  \\ 
\midrule
NSR = 0.2               & 8/8    & 100\% \\ 
NSR = 0.5               & 7/8    & 88\%  \\
NSR = 1                 & 4/8    & 50\%  \\ 
\midrule
INF=5                   & 8/12   & 67\%  \\ 
INF=20                  & 11/12  & 92\%  \\ 
\midrule
k=100                   & 11/12  & 92\%  \\ 
k=500                   & 8/12   & 67\%  \\
\bottomrule
\end{tabular*}
\item[$^*$] Superior performance is measured by means of the median {FPR}.
\end{table*}
\end{center}

\begin{figure}[ht!]
    \centering
    \includegraphics{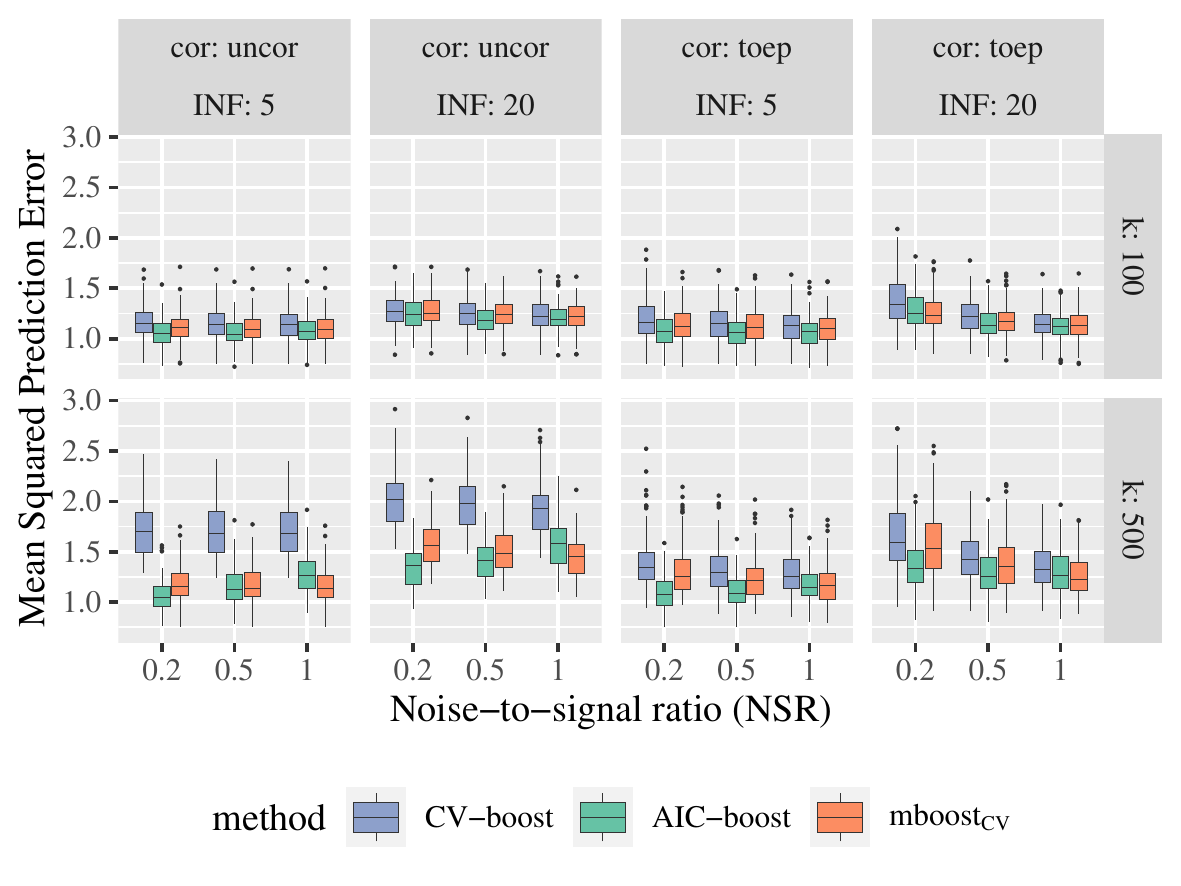}
    \caption{Mean Squared Prediction Error for new variable selection strategies by simulation setting.}
    \label{fig: MSPE full}
\end{figure}

\begin{figure}[ht!]
    \centering
    \includegraphics{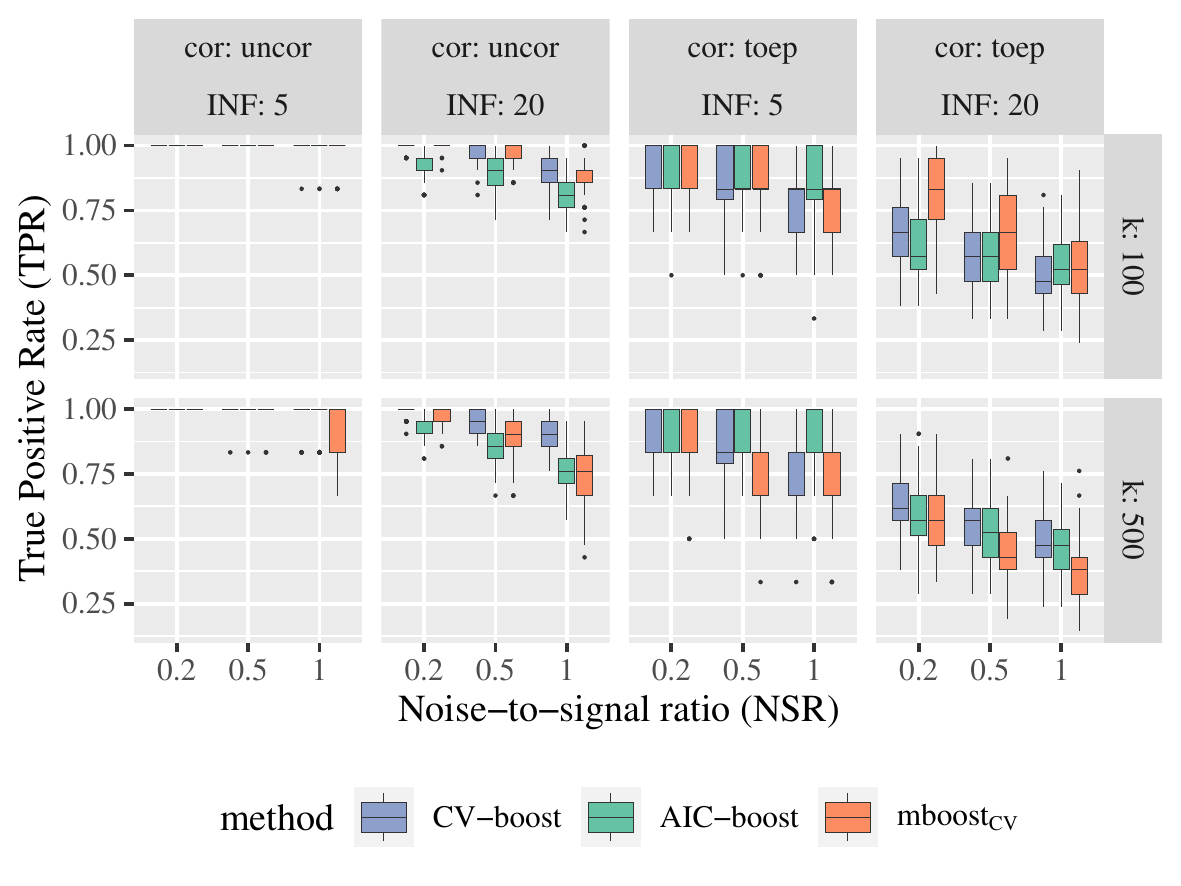}
    \caption{True Positive Rate for new variable selection strategies by simulation setting.}
    \label{fig: TPR}
\end{figure}

\begin{figure}[t!]
    \centering
    \includegraphics{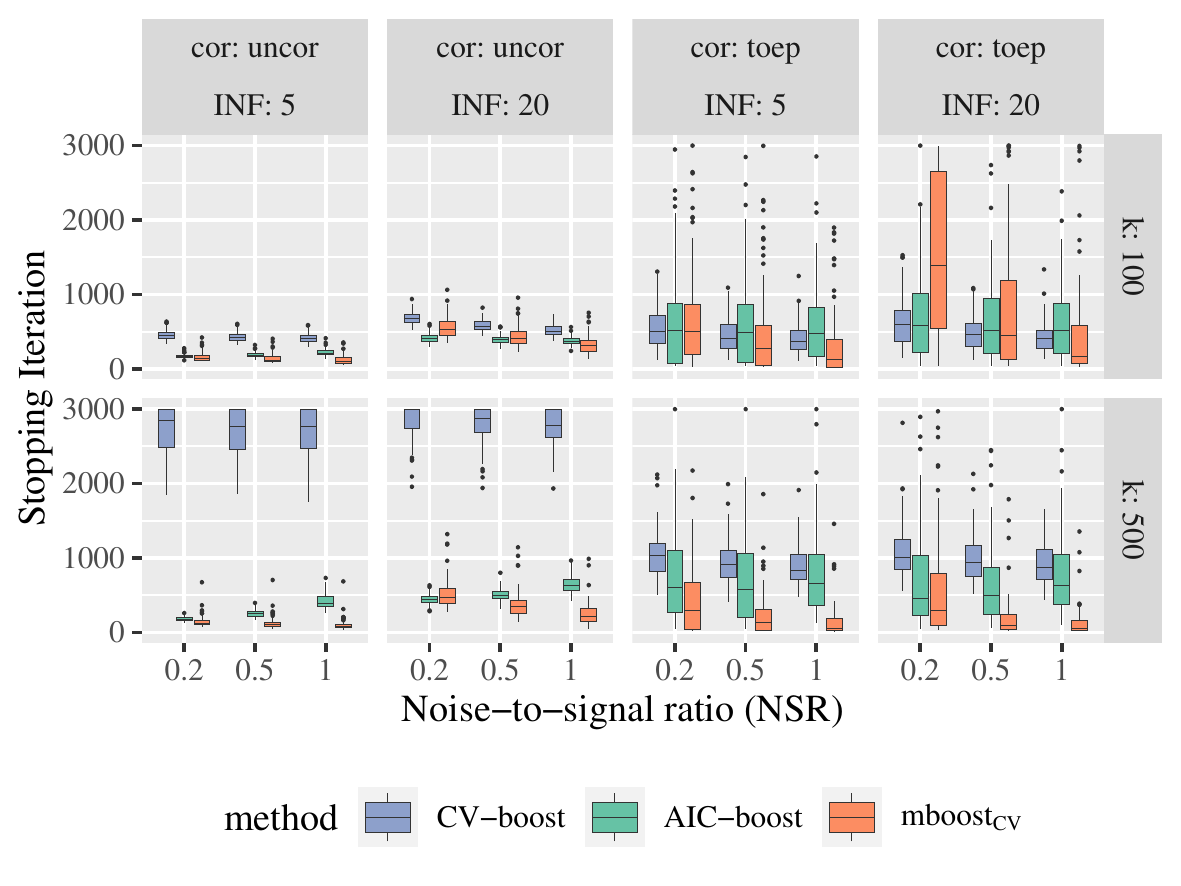}
    \caption{Stopping iteration $t^*$ for new variable selection strategies by simulation setting.}
    \label{fig: SI}
\end{figure}

\clearpage

\subsection{Simulation Results of Poisson Distributed Data}
\label{append sim_pois}

\begin{figure}[ht!]
    \centering
    \includegraphics{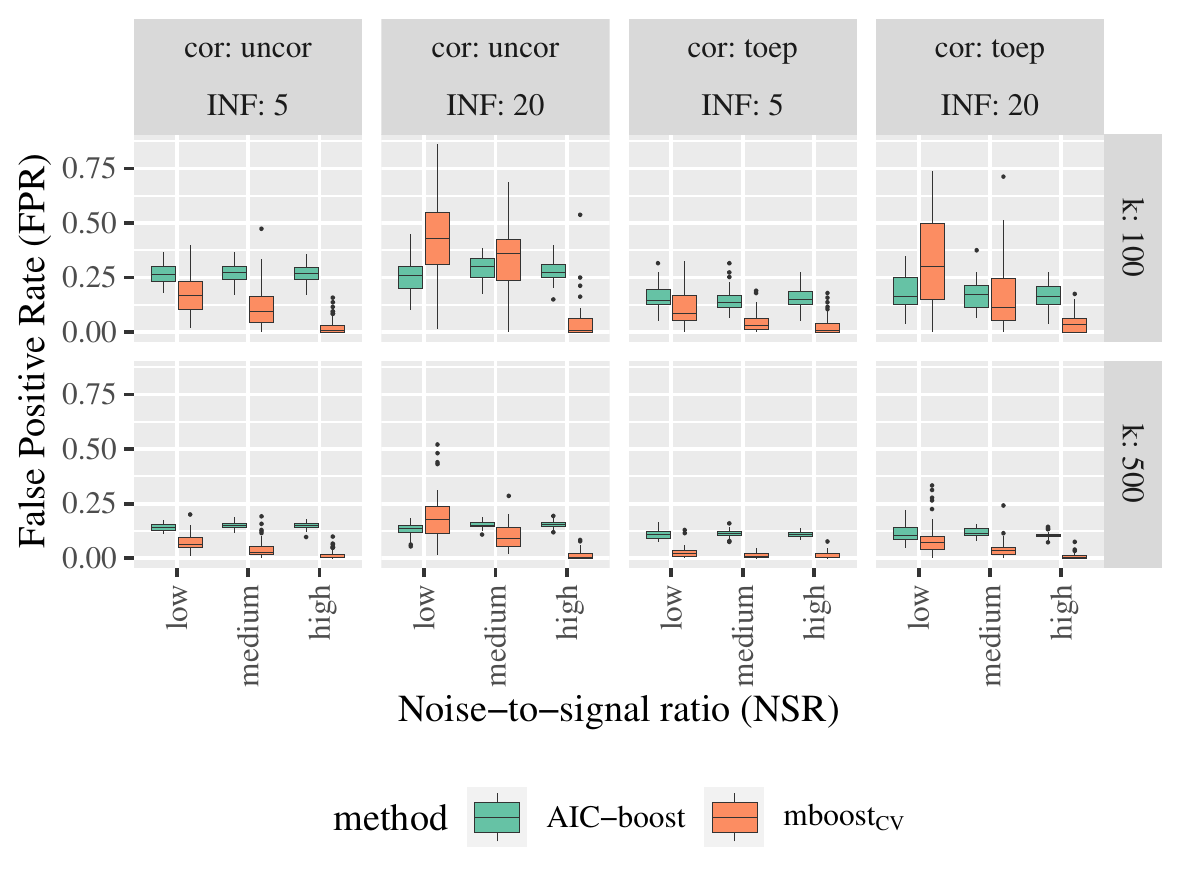}
    \caption{False Positive Rate for new variable selection strategies by simulation setting.}
    \label{fig: FPR full pois}
\end{figure}

\begin{figure}[ht!]
    \centering
    \includegraphics{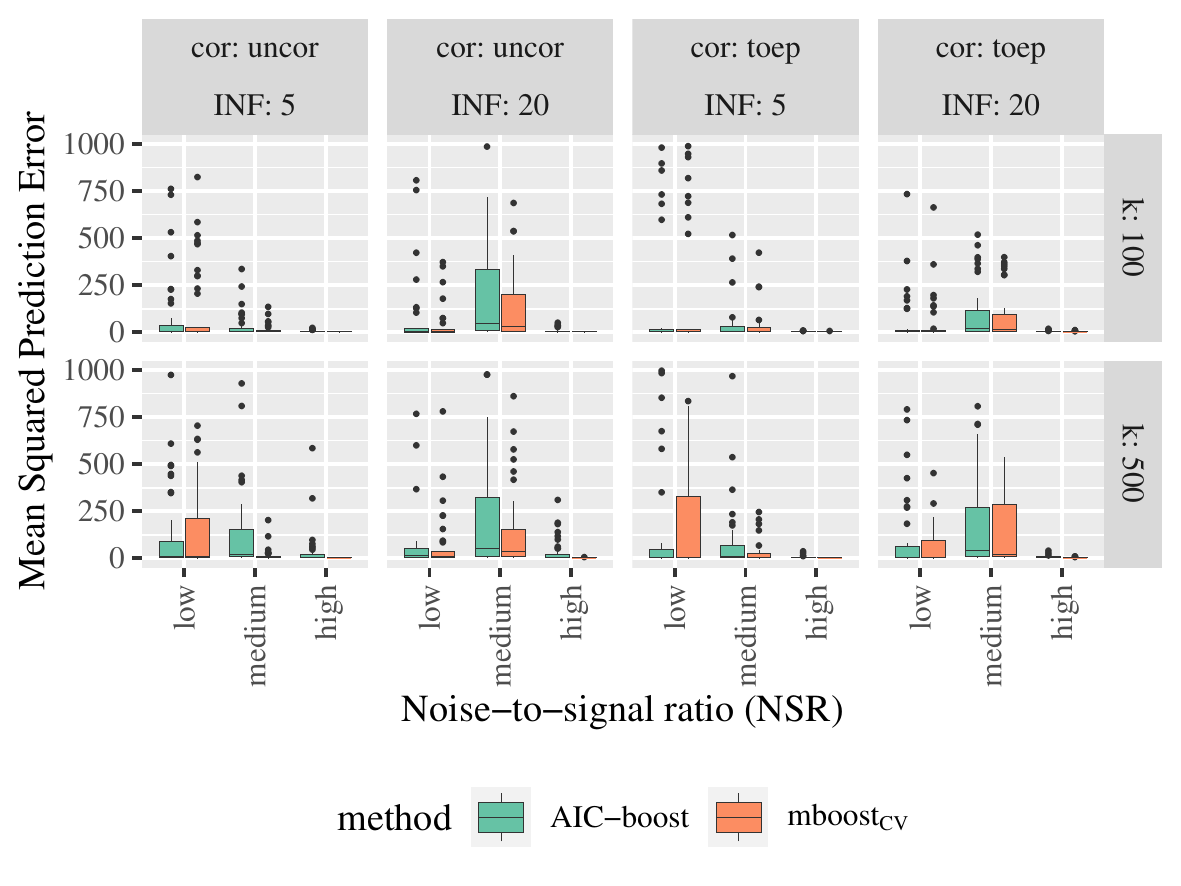}
    \caption{Mean Squared Prediction Error for new variable selection strategies by simulation setting.}
    \label{fig: MSPE full pois}
\end{figure}

\begin{figure}[ht!]
    \centering
    \includegraphics{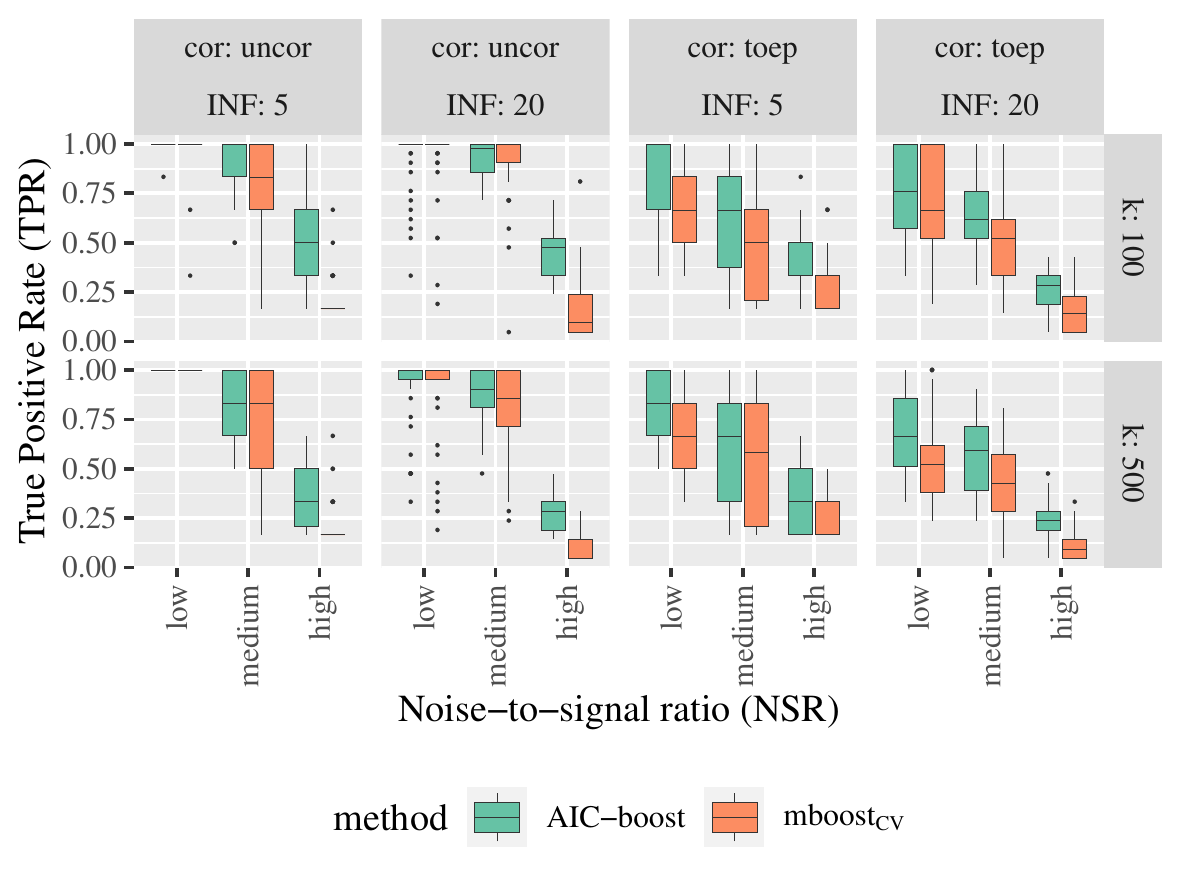}
    \caption{True Positive Rate for new variable selection strategies by simulation setting.}
    \label{fig: TPR pois}
\end{figure}

\begin{figure}[t!]
    \centering
    \includegraphics{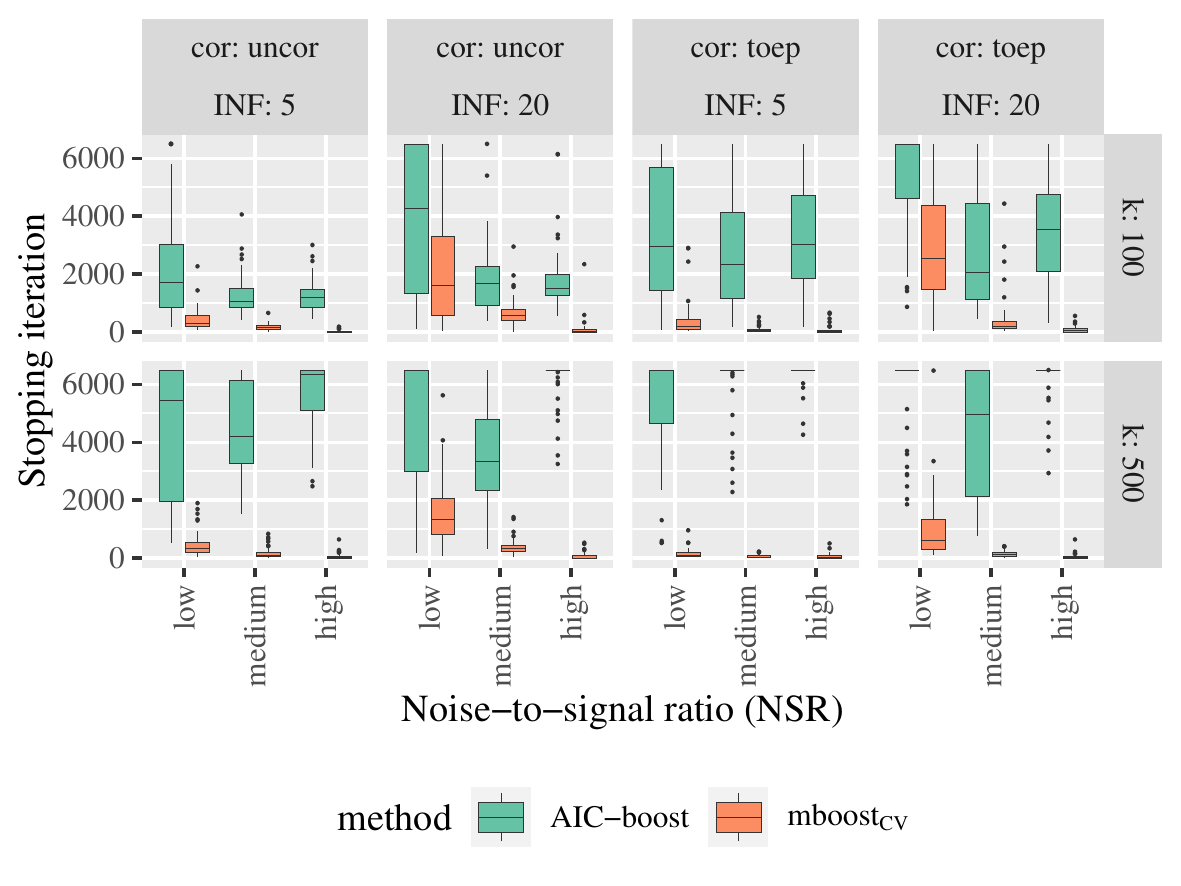}
    \caption{Stopping iteration $t^*$ for new variable selection strategies by simulation setting.}
    \label{fig: SI pois}
\end{figure}

\clearpage
\section{Application Results \label{app1}}

\subsection{Model Table of Application}

\begin{sidewaystable}
\caption{Coefficients of standardized covariates on log(age-standardized incidence rate) of different estimation models. Numbers in brackets correspond to the ordering of the absolute values of the coefficients.\label{tab: coef covid}}%
\begin{tabular*}{\textheight}{@{\extracolsep\fill}l D{.}{.}{4}c D{.}{.}{4}c D{.}{.}{4}c@{\extracolsep\fill}}
\toprule
\multicolumn{1}{@{}l@{}}{\textbf{Covariate}} &\multicolumn{2}{@{}l@{}}{\textbf{$\text{mboost}_{\text{CV}}$}}
&\multicolumn{2}{@{}c@{}}{\textbf{LASSO}} &\multicolumn{2}{@{}c@{}}{\textbf{AIC-boost}}  \\ 
\midrule
Intercept & 4.133&     & 3.992&      & 2.829   &       \\
Age-standardized incidence rate per 100.000 person-years until 15.03.                           & 0.043 & (5) & 0.046  & (6)  & 0.109     & (6)   \\
Unemployment rate of young persons (under   26 years) in 2017                                     & -0.071                    & (4) & -0.055 & (5)  & -0.071    & (8)   \\
\%Voter turnout (Number of valid votes in   the last Bundestag election) of all registered voters & 0.023                     & (7) & 0.036  & (7)  & 0.033     & (13)  \\
\%Roman-catholics in 2011                                                                         & 0.075                     & (3) & 0.079  & (4)  & 0.086     & (7)   \\
Persons in long-term care per 10.000   persons in 2017                                            & -0.024                    & (6) & -0.033 & (8)  & -0.039    & (12)  \\
Premature mortality (deaths of persons   younger than 65 years) per 1,000 persons                 & -0.097                    & (2) & -0.088 & (3)  & -0.117    & (5)   \\
\%Households with low income (1,500€   per month) in all households in 2016                       & -0.015                    & (8) & -0.015 & (9)  & -0.012    & (18)  \\
Latitude                                                                                          & -0.122                    & (1) & -0.130 & (2)  & -0.130    & (4)   \\
\%Older employed persons (55 years+) in   all employed persons in 2017                            &                           &     & -0.013 & (10) &           &       \\
Ever 100+ inbound commuters from   Tirschenreuth                                                  &                           &     & 0.141  & (1)  & 0.920     & (1)   \\
\%Change of number of persons at age 50-65   in 2012-2017                                         &                           &     &        &      & 0.009     & (19)  \\
Sex ratio (females to males) at age 20-40   in 2017                                               &                           &     &        &      & 0.021     & (17)  \\
Total sex ratio (females to males) in   2017                                                      &                           &     &        &      & 0.068     & (9)   \\
\%Young employed persons in all young   persons (under 26 years) in 2017                          &                           &     &        &      & 0.054     & (11)  \\
\%Older employed persons in all older   persons (55 years+) in 2011-2017                          &                           &     &        &      & 0.027     & (15)  \\
\%Change of number of employed persons in   2012-2016                                             &                           &     &        &      & 0.009     & (19)  \\
Average travel time to the next   large-sized regional center ("Oberzentrum")                      &                           &     &        &      & -0.032    & (14)  \\
\%Outbound commuters over a distance of   50km+ in all employed persons in 2017                   &                           &     &        &      & -0.006    & (20)  \\
\%Outbound commuters over a distance of   150km+ in all employed persons in 2017                  &                           &     &        &      & -0.056    & (10)  \\
Ever 100+ inbound commuters from   Hohenlohekreis                                                 &                           &     &        &      & 0.136     & (3)   \\
Ever 100+ inbound commuters from Aachen                                                           &                           &     &        &      & 0.027     & (15)  \\
Ever 100+ inbound commuters from   Rosenheim (city + district)                                    &                           &     &        &      & 0.194     & (2)   \\
\bottomrule
\end{tabular*}
\end{sidewaystable}

\clearpage

\bibliographystyle{unsrt}  
\bibliography{references}

\begin{thebibliography}{10}

\bibitem{Buhlmann.2007}
Peter B{\"u}hlmann and Torsten Hothorn.
\newblock {B}oosting {A}lgorithms: {R}egularization, {P}rediction and {M}odel
  {F}itting.
\newblock {\em Statistical Science}, 22(4):477--505, 2007.

\bibitem{Freund.1996}
Yoav Freund and Robert~E Schapire.
\newblock Experiments with a new boosting algorithm.
\newblock {\em Proceedings of the Thirteenth International Conference on
  Machine Learning Theory}, pages 148--156, 1996.

\bibitem{breiman1997arcing}
Leo Breiman.
\newblock {A}rcing the edge.
\newblock Technical report, 486, Statistics Department, University of
  California at~Berkeley, 1997.

\bibitem{Friedman.2000}
Jerome Friedman, Trevor Hastie, and Robert Tibshirani.
\newblock Additive logistic regression: a statistical view of boosting (with
  discussion and a rejoinder by the authors).
\newblock {\em The Annals of Statistics}, 28(2):337--407, 2000.

\bibitem{Friedman.2001}
Jerome Friedman.
\newblock Greedy function approximation: A gradient boosting machine.
\newblock {\em The Annals of Statistics}, 29(5):1189--1232, 2001.

\bibitem{Mayr.2012}
Andreas Mayr, Benjamin Hofner, and Matthias Schmid.
\newblock {T}he {I}mportance of {K}nowing {W}hen to {S}top.
\newblock {\em Methods of Information in Medicine}, 51(02):178--186, 2012.

\bibitem{Thomas.2017}
Janek Thomas, Tobias Hepp, Andreas Mayr, Bernd Bischl, and Yuhai Zhao.
\newblock {P}robing for {S}parse and {F}ast {V}ariable {S}election with
  {M}odel-{B}ased {B}oosting.
\newblock {\em Computational and Mathematical Methods in Medicine}, 2017:1--8,
  2017.

\bibitem{meinshausen2010stability}
Nicolai Meinshausen and Peter B{\"u}hlmann.
\newblock {S}tability selection.
\newblock {\em Journal of the Royal Statistical Society: Series B (Statistical
  Methodology)}, 72(4):417--473, 2010.

\bibitem{hofner2015controlling}
Benjamin Hofner, Luigi Boccuto, and Markus G{\"o}ker.
\newblock {C}ontrolling false discoveries in high-dimensional situations:
  {B}oosting with stability selection.
\newblock {\em BMC bioinformatics}, 16(1):1--17, 2015.

\bibitem{stromer2021deselection}
Annika Str{\"o}mer, Christian Staerk, Nadja Klein, Leonie Weinhold, Stephanie
  Titze, and Andreas Mayr.
\newblock {D}eselection of base-learners for statistical boosting—with an
  application to distributional regression.
\newblock {\em Statistical Methods in Medical Research}, 31(2):207--224, 2022.

\bibitem{buhlmann2010twin}
Peter B{\"u}hlmann and Torsten Hothorn.
\newblock {T}win boosting: improved feature selection and prediction.
\newblock {\em Statistics and Computing}, 20(2):119--138, 2010.

\bibitem{staerk2021subspace}
Christian Staerk and Andreas Mayr.
\newblock {R}andomized boosting with multivariable base-learners for
  high-dimensional variable selection and prediction.
\newblock {\em BMC bioinformatics}, 22(1):1--28, 2021.

\bibitem{buhlmann2006sparse}
Peter B{\"u}hlmann, Bin Yu, Yoram Singer, and Larry Wasserman.
\newblock Sparse boosting.
\newblock {\em Journal of Machine Learning Research}, 7(6):1001--1024, 2006.

\bibitem{Hofner.2011}
Benjamin Hofner, Torsten Hothorn, Thomas Kneib, and Matthias Schmid.
\newblock {A} {F}ramework for {U}nbiased {M}odel {S}election {B}ased on
  {B}oosting.
\newblock {\em Journal of Computational and Graphical Statistics},
  20(4):956--971, 2011.

\bibitem{tutz2010generalized}
Gerhard Tutz and Andreas Groll.
\newblock {G}eneralized {L}inear {M}ixed {M}odels {B}ased on {B}oosting.
\newblock {\em Statistical Modelling and Regression Structures: Festschrift in
  Honour of Ludwig Fahrmeir}, pages 197--215, 2010.

\bibitem{Fahrmeir2021glm}
Ludwig Fahrmeir, Thomas Kneib, Stefan Lang, and Brian~D. Marx.
\newblock Generalized linear models.
\newblock {\em Regression Models}, pages 283--342, 2021.

\bibitem{Hastie2007}
Trevor Hastie.
\newblock {C}omment: {B}oosting {A}lgorithms: {R}egularization, {P}rediction
  and {M}odel {F}itting.
\newblock {\em Statistical Science}, 22(4):513--515, November 2007.

\bibitem{mboost2}
Torsten Hothorn, Peter B{{\"u}}hlmann, Thomas Kneib, Matthias Schmid, and
  Benjamin Hofner.
\newblock {M}odel-based {B}oosting 2.0.
\newblock {\em Journal of Machine Learning Research}, 11(71):2109--2113, 2010.

\bibitem{Hepp.2016}
Tobias Hepp, Matthias Schmid, Olaf Gefeller, Elisabeth Waldmann, and Andreas
  Mayr.
\newblock {A}pproaches to {R}egularized {R}egression - {A} {C}omparison between
  {G}radient {B}oosting and the {L}asso.
\newblock {\em Methods of Information in Medicine}, 55(5):422--430, 2016.

\bibitem{Doblhammer2022}
Gabriele Doblhammer, Constantin Reinke, and Daniel Kreft.
\newblock Social disparities in the first wave of {COVID}-19 incidence rates in
  {G}ermany: a county-scale explainable machine learning approach.
\newblock {\em {BMJ} Open}, 12(2: e049852):1--11, February 2022.

\bibitem{Plumper.2020}
Thomas Pl{\"u}mper and Eric Neumayer.
\newblock {T}he pandemic predominantly hits poor neighbourhoods? {SARS-C}o{V}-2
  infections and {COVID-19} fatalities in {G}erman districts.
\newblock {\em {E}uropean journal of public health}, 30(6):1176--1180, 2020.

\bibitem{Wachtler.2020}
Benjamin Wachtler, Niels Michalski, Enno Nowossadeck, Michaela Diercke, Morten
  Wahrendorf, Claudia Santos-H{\"o}vener, Thomas Lampert, and Jens Hoebel.
\newblock {S}ocioeconomic inequalities in the risk of {SARS}-{C}o{V}-2
  infection - {F}irst results from an analysis of surveillance data from
  {G}ermany.
\newblock {\em Journal of health monitoring}, 5(Suppl 7):18--29, 2020.

\bibitem{Rohleder.2022}
Sven Rohleder, Diogo Costa, and Kayvan Bozorgmehr.
\newblock {A}rea-level socioeconomic deprivation, non-national residency, and
  {C}ovid-19 incidence: {A} longitudinal spatiotemporal analysis in {G}ermany.
\newblock {\em EClinicalMedicine}, 49:101485, 2022.

\bibitem{Plumper.2018}
Thomas Pl{\"u}mper, Denise Laroze, and Eric Neumayer.
\newblock {T}he limits to equivalent living conditions: regional disparities in
  premature mortality in {G}ermany.
\newblock {\em {J}ournal of public health}, 26(3):309--319, 2018.

\bibitem{Brandl.2020}
M.~Brandl, R.~Selb, S.~Seidl-Pillmeier, D.~Marosevic, U.~Buchholz, and
  S.~Rehmet.
\newblock {M}ass gathering events and undetected transmission of {SARS-C}o{V}-2
  in vulnerable populations leading to an outbreak with high case fatality
  ratio in the district of {T}irschenreuth, {G}ermany.
\newblock {\em {E}pidemiology and infection}, 148:e252, 2020.

\bibitem{Fuest.2019}
Clemens Fuest and Lea Immel.
\newblock {E}in zunehmend gespaltenes {L}and? -- {R}egionale
  {E}inkommensunterschiede und die {E}ntwicklung des {G}ef{\"a}lles zwischen
  {S}tadt und {L}and sowie {W}est- und {O}stdeutschland.
\newblock {\em ifo Schnelldienst}, 72(16):19--28, 2019.

\bibitem{Ballering.2021}
Aranka~Vivi{\"e}nne Ballering, Sabine Oertelt-Prigione, Tim~C. {Olde Hartman},
  and Judith G.~M. Rosmalen.
\newblock {S}ex and {G}ender-{R}elated {D}ifferences in {COVID-19} {D}iagnoses
  and {SARS-C}o{V}-2 {T}esting {P}ractices {D}uring the {F}irst {W}ave of the
  {P}andemic: {T}he {D}utch {L}ifelines {COVID-19} {C}ohort {S}tudy.
\newblock {\em Journal of women's health}, 30(12):1686--1692, 2021.

\bibitem{Bianconi.2020}
Vanessa Bianconi, Massimo~R. Mannarino, Paola Bronzo, Ettore Marini, and Matteo
  Pirro.
\newblock {T}ime-related changes in sex distribution of {COVID-19} incidence
  proportion in {I}taly.
\newblock {\em Heliyon}, 6(10):e05304, 2020.

\bibitem{Doerre.2022}
Achim Doerre and Gabriele Doblhammer.
\newblock The influence of gender on {COVID-19} infections and mortality in
  {G}ermany: Insights from age- and gender-specific modeling of contact rates,
  infections, and deaths in the early phase of the pandemic.
\newblock {\em PloS one}, 17(5):e0268119, 2022.

\bibitem{Ancochea.2021}
Julio Ancochea, Jose~L. Izquierdo, and Joan~B. Soriano.
\newblock {E}vidence of {G}ender {D}ifferences in the {D}iagnosis and
  {M}anagement of {C}oronavirus {D}isease 2019 {P}atients: {A}n {A}nalysis of
  {E}lectronic {H}ealth {R}ecords {U}sing {N}atural {L}anguage {P}rocessing and
  {M}achine {L}earning.
\newblock {\em Journal of women's health}, 30(3):393--404, 2021.

\bibitem{Leibert.2022}
Tim Leibert, Manuel Wolff, and Annegret Haase.
\newblock {S}hifting spatial patterns in {G}erman population trends:
  local-level hot and cold spots, 1990--2019.
\newblock {\em Geographica Helvetica}, 77(3):369--387, 2022.

\bibitem{Fink.2019}
Philipp Fink, Martin Hennicke, and Heinrich Tiemann.
\newblock {\em {U}nequal {G}ermany: {S}ocio-economic disparities report 2019}.
\newblock For a better tomorrow. Friedrich-Ebert-Stiftung, Bonn, 2019.

\bibitem{Robinson1950}
W.~S. Robinson.
\newblock {E}cological {C}orrelations and the {B}ehavior of {I}ndividuals.
\newblock {\em American Sociological Review}, 15(3):351--357, June 1950.

\bibitem{hurvich1989regression}
Clifford~M Hurvich and Chih-Ling Tsai.
\newblock {R}egression and time series model selection in small samples.
\newblock {\em Biometrika}, 76(2):297--307, 1989.

\bibitem{Buehlmann2007rejoinder}
Peter B{\"u}hlmann and Torsten Hothorn.
\newblock {R}ejoinder: {B}oosting {A}lgorithms: {R}egularization, {P}rediction
  and {M}odel {F}itting.
\newblock {\em Statistical Science}, 22(4):516--522, 2007.

\end{thebibliography}

\end{document}